\documentclass[aps,pra,superscriptaddress,twocolumn,10pt]{revtex4-2}

\usepackage[utf8]{inputenc}
\usepackage[T1]{fontenc}
\usepackage{quantikz}
\usepackage{amsmath, mleftright}
\usepackage{mathtools}
\usepackage{braket}
\usepackage{amsfonts}
\usepackage{amssymb}
\usepackage{lipsum}
\usepackage{dsfont}
\usepackage{stmaryrd}
\usepackage{graphicx}
\usepackage{booktabs}
\usepackage{multirow}
\usepackage{siunitx}
\usepackage[version=4]{mhchem}
\usepackage{chemfig}
\usepackage[hidelinks]{hyperref}
\usepackage[capitalise,noabbrev]{cleveref}
\usepackage{xcolor}
\usepackage{microtype}
\usepackage{algorithm2e}
\usepackage{algpseudocode}
\usepackage{acro}
\usepackage[normalem]{ulem}
\usepackage{makecell}
\usepackage{bbold}
\usepackage{listings,xfp}
\usepackage{anyfontsize}

\renewcommand{\dag}{^\dagger}

\DeclarePairedDelimiter\ceil{\lceil}{\rceil}

\DeclarePairedDelimiter\norm{\lVert}{\rVert}

\newcommand{\id}{\mathbb{1}}
\newcommand{\R}{\ensuremath{\mathbb{R}}}

\renewcommand{\lg}{\log_2}

\newcommand{\crea}[2]{a^{#1\mspace{-0.5mu}\raisebox{0.3ex}{$\scriptstyle\dagger$}}_{\mspace{-0mu}#2^{#1}}}

\newcommand{\anni}[2]{a^{#1}_{\mspace{-0mu}#2^{#1}}}

\newcommand{\approptoinn}[2]{\mathrel{\vcenter{
    \offinterlineskip\halign{\hfil$##$\cr
    #1\propto\cr\noalign{\kern2pt}#1\sim\cr\noalign{\kern-2pt}}}}}

\newcommand{\defeq}{\vcentcolon=}

\DeclareMathOperator{\node}{nd} % graph node
\DeclareSIUnit\hartree{\text {\ensuremath {E}}_{\mathrm {h}}}

\makeatletter
{\large % Capture font definitions of \large
\xdef\f@size@large{\f@size}
\xdef\f@baselineskip@large{\f@baselineskip}
\normalsize % Capture font definitions for \normalsize
\xdef\f@size@normalsize{\f@size}
\xdef\f@baselineskip@normalsize{\f@baselineskip}
}
\newcommand{\mediumlarge}{%
\fontsize
{\fpeval{(2*\f@size@large+\f@size@normalsize)/3}}
{\fpeval{(2*\f@baselineskip@large+\f@baselineskip@normalsize)/3}}%
\selectfont
}
\makeatother

\usetikzlibrary{positioning}
\usetikzlibrary{shapes.geometric}
\usetikzlibrary{shapes.misc}

\newcommand{\stkout}[1]{\ifmmode\text{\sout{\ensuremath{#1}}}\else\sout{#1}\fi}

\def\conflictsname{Conflicts of interest}
\newenvironment{conflicts}{%
    \expandafter\section\expandafter*\expandafter{\conflictsname}%
    }%

\def\dataavailname{Data availability}
\newenvironment{dataav}{%
    \expandafter\section\expandafter*\expandafter{\dataavailname}%
    }%

\newcommand{\auchem}{Department of Chemistry, Aarhus University, DK-8000 Aarhus C, Denmark}
\newcommand{\auphys}{Department of Physics and Astronomy, Aarhus University, DK-8000 Aarhus C, Denmark}
\newcommand{\aucs}{Department of Computer Science, Aarhus University, DK-8200 Aarhus N, Denmark}
\newcommand{\kvantify}{Kvantify Aps, DK-2300 Copenhagen S, Denmark}

\DeclareAcronym{hf}{
    short = HF ,
    long = Hartree-Fock ,
}
\DeclareAcronym{vscf}{
    short = VSCF,
    long = Vibrational Self-Consistent Field
}
\DeclareAcronym{pes}{
    short = PES,
    long = Potential Energy Surface
}
\DeclareAcronym{keo}{
    short = KEO,
    long = Kinetic Energy Operator
}
\DeclareAcronym{fvci}{
short = FVCI,
long = Full Vibrational Configuration Interaction
}
\DeclareAcronym{vcc}{
short = VCC,
long = Vibrational Coupled Cluster
}
\DeclareAcronym{cp}{
short = CP,
long = canonical polyadic
}
\DeclareAcronym{qpe}{
short = QPE,
long = quantum phase estimation
}
\DeclareAcronym{sop}{
short = SOP,
long = sum-over-product
}
\DeclareAcronym{lcu}{
short = LCU,
long = linear combination of unitaries
}
\DeclareAcronym{svd}{
short = SVD,
long = singular value decomposition
}
\DeclareAcronym{hooi}{
short = HOOI,
long = higher order orthogonal iteration
}
\DeclareAcronym{cpals}{
short = CP-ALS,
long = CANDECOMP/PARAFAC alternating least square
}
\DeclareAcronym{qft}{
short = QFT,
long = Quantum Fourier Transform
}
\DeclareAcronym{pah}{
short = PAH,
long = polycyclic aromatic hydrocarbon
}
\DeclareAcronym{adga}{
short = ADGA ,
long = adaptive density-guided approach ,
}
\DeclareAcronym{holc}{
short = HOLC ,
long = hybrid optimized and localized coordinate ,
}

\begin{document}
    \title{Fault-tolerant quantum computations of vibrational wave functions}

    \author{Marco Majland}
    \affiliation{\kvantify}
    \affiliation{\auphys}
    \affiliation{\auchem}

    \author{Rasmus Berg Jensen}
    \affiliation{\kvantify}
%\affiliation{\auphys}
    \affiliation{\auchem}

    \author{Patrick Ettenhuber}
    \affiliation{\kvantify}

    \author{Irfansha Shaik}
    \affiliation{\kvantify}
    \affiliation{\aucs}

    \author{Nikolaj Thomas Zinner}
    \affiliation{\kvantify}
    \affiliation{\auphys}

    \author{Ove Christiansen}
    \affiliation{\kvantify}
    \affiliation{\auchem}

    \begin{abstract}
        Quantum computation of vibrational properties of molecules is a promising platform to obtain computational advantages for computational chemistry. However, fault-tolerant quantum computations of vibrational properties remain a relatively unexplored field in quantum computing. In this work, we present different algorithms for efficient encodings of vibrational Hamiltonians using qubitization. Specifically, we investigate different encoding representations, high order tensor decomposition to obtain low rank approximations for the vibrational Hamiltonian, rectilinear and polyspherical coordinate systems, parallelization and grouping algorithms. To investigate the performance of the different methods, we perform benchmark computations for both small and large molecules with more than one hundred vibrational modes.
    \end{abstract}

    \maketitle

    %\tableofcontents

    \section{Introduction}\label{sec:introduction}

    The accurate computation of the quantum states of molecules, which comprise many interacting electrons and nuclei, remains a formidable challenge, even for the most powerful high-performance computers of today. Quantum simulation of such many-body systems offers a promising path forward, with quantum chemistry widely considered one of the fields where quantum algorithms may provide substantial computational advantages~\cite{daley_practical_2022,lee_is_2023,elfving_how_2020,aspuru-guzik_simulated_2005}. In this context, the \ac{qpe} algorithm formally presents a powerful framework for achieving quantum speedups in the calculation of molecular properties, such as ground and excited state energies and other observables~\cite{aspuru-guzik_simulated_2005,knill_optimal_2007,rall_quantum_2020,steudtner_fault-tolerant_2023,obrien_efficient_2022,abrams_quantum_1999,reiher_elucidating_2017}. 
    
    To date, most quantum computational efforts have concentrated on the electronic structure problem, namely, computing electronic energies and wave functions. Although these are central and logical starting points, as the electronic energies define the \ac{pes} when viewed as functions of nuclear positions, a complete molecular description must also account for the quantum states due to nuclear motion. These nuclear motions contribute critically to molecular free energies, key drivers of chemical reactivity and stability, and are often observed directly in spectroscopy.
    
    In recent years, studies have begun to explore the calculation of molecular vibrational properties using quantum computers, employing both Noisy Intermediate Scale Quantum (NISQ) devices and fault-tolerant quantum computing approaches~\cite{mcardle_digital_2019,ollitrault_hardware_2020,sawaya_near-_2021,sawaya_resource-efficient_2020,sparrow_simulating_2018,magann_digital_2021,richerme_quantum_2023,sawaya_quantum_2019,jahangiri_quantum_2020}. 
    Variational algorithms based on variants of unitary vibrational coupled cluster theory have been proposed to variationally encode vibrational wavefunctions for NISQ devices~\cite{mcardle_digital_2019,ollitrault_hardware_2020,majland_optimizing_2023}. 
    In the context of fault-tolerant algorithms primarily Trotter methods being investigated. ~\cite{sawaya_resource-efficient_2020,trenev_refining_2023}. Using rough estimates for several vibrational Hamiltonians, is has been argued that 
    the cost of simulating vibrational properties may be tractable earlier compared to its electronic counterpart~\cite{sawaya_near-_2021}. As part of such further speculations lower bounds and gate estimates of Trotter steps for acetylene-like polyynes were considered in Ref.~\cite{trenev_refining_2023} using the methods presented in Ref.~\cite{childs_theory_2021}.
    However, in the field of electronic structure, qubitization has been shown to require fewer Toffoli gates for practical calculations compared to other methods~\cite{babbush_encoding_2018,lee_even_2021,berry_qubitization_2019, kivlichan_quantum_2018,babbush_low-depth_2018}. 
    Qubitization encodes the Hamiltonian as a Szegedy walk operator which may be implemented with zero error excluding the error arising from rotation synthesis~\cite{low_hamiltonian_2019,babbush_encoding_2018}. In contrast, Trotter methods implement the time evolution operator of the Hamiltonian which yield finite errors in the implementation~\cite{kivlichan_quantum_2018,babbush_low-depth_2018,childs_theory_2021}. Since current encodings for vibrational wave functions rely on Trotter methods, any potential advantages of utilizing qubitization for encoding the vibrational Hamiltonian remains to be investigated.
    Furthermore, although substantial progress has been made in resource reduction techniques for electronic structure computations, only a limited number of optimization strategies have been suggested to minimize computational overhead in fault-tolerant vibrational computations~\cite{sawaya_resource-efficient_2020,trenev_refining_2023}.
    We here note the recent interesting work on Trotter simulation with cost estimates that employs resource reduction techniques both similar and different to ours.~\cite{malpathak2025trottersimulationvibrationalhamiltonians}

    Cost estimates for vibrational computations necessitate consideration of vibrational Hamiltonians, including coordinate selection, \ac{keo} representation, and the \ac{pes}. Past research often focused on simple vibrational potentials describing small systems, using only rectilinear coordinates like normal coordinates. These are inadequate for large amplitude nuclear motions, such as torsional motions crucial in many biomolecular systems. Choosing non-optimal coordinates can lead to strong inter-coordinate couplings. Therefore, employing curvilinear coordinates for describing molecular internal motion, which naturally include angles and bonds, can be beneficial for reducing \ac{pes} and wave function couplings.

    In this work, we present an algorithm for encoding the vibrational Hamiltonian using qubitization. Detailed quantum circuits for block encodings of vibrational Hamiltonians are discussed. In order to optimize relevant parameters for qubitization, we discuss different representations for encoding vibrational second quantization operators. To reduce gate costs of block encoding when using chemically relevant many-term Hamiltonians, we present tensor decomposition methods utilizing a combination of the Tucker and canonical decompositions of multidimensional tensors which allow for efficient low-rank approximations of the vibrational Hamiltonian. In addition, we present efficient grouping algorithms to perform parallel block encodings which further reduce the circuit depths.

     To illustrate the applicability of our method to general curvilinear coordinate systems, we apply it to hydrogen peroxide using polyspherical coordinates. Leveraging recent techniques~\cite{bader_efficient_2024}, we obtain an efficient representation of the vibrational \ac{keo} with low coupling complexity and manageable computational cost. Similar to how the \ac{pes} is often approximated in practice, the \ac{keo} can also be simplified. By following general frameworks for constructing exact \ac{keo}s~\cite{lauvergnat_exact_2002}, one can systematically derive approximate forms that retain accuracy while drastically reducing the number of coupled terms~\cite{bader_efficient_2024}.
    
    To investigate the scaling of larger systems, we provide gate costs for encoding the vibrational Hamiltonian for large molecules ranging from 10 to 100 vibrational modes to perform realistic estimates of the required quantum resources for performing vibrational wave function calculations on fault-tolerant quantum computers.

    \section{Theoretical background}
    \label{sec:theoretical_background}
    
    In this section, we discuss the theoretical background relating to working conditions for the algorithms and computations discussed later. Specifically, we discuss many-mode second quantization formalism for vibrational wave function theory, the choice of coordinates and \ac{keo}s, the representation of the full Hamiltonian, and finally qubitization. 
    \subsection{Many-mode second quantization}
    \label{sec:vibrational_structure}
    We shall from the outset use many-mode second quantization as introduced in Ref.~\cite{christiansen_second_2004}. Let $\{m_1,m_2,...\}$ denote distinguishable vibrational modes and let $\{p^m,q^m,r^m,s^m\}$ index general one-mode functions (modals) for a mode $m$. We use $N_m$ to denote the number of modals for mode $m$.  We define creation and annihilation operators $a^{m_1\dagger}_{p^{m_1}}$ and $a^{m_2}_{q^{m_2}}$ which create and annihilate occupation of modals with indices $p^{m_1}$ and $q^{m_2}$ for two given modes $m_1$ and $m_2$ respectively. The space of occupation vectors defining occupation for all mode and modals is larger than the relevant physical space defined by $\{\forall m: \sum_{p^m} k_{r^m}^m= 1\}$ where $k_{r^m}^m$ is the occupation of the modal $p^m$ for the mode $m$. The commutation relations for the creation and annihilation operators may be written as
    \begin{align}
    \label{eq:commutator_relation1}
        [a_{p^{m_1}}^{m_1}, a^{m_2\dagger}_{q^{m_2}}] = \delta_{p^{m_1}q^{m_2}}\delta_{m_1m_2}
        \intertext{and}
       [a^{m_1\dagger}_{p^{1_2}}, a^{m_2\dagger}_{q^{m_2}}] = [a_{p^{m_1}}^{m_1}, a_{q^{m_2}}^{m_2}] = 0 \label{eq:commutator_relation2}
   \end{align}
    Importantly, the creation and annihilation operators are not to be confused with harmonic oscillator ladder operators. The modal functions are not assumed to be harmonic oscillator functions but are general one-mode basis functions. In practical calculations, these modals may be obtained from \ac{vscf} theory.
    Recalling that VSCF is exact for uncoupled anharmonic oscillators, the VSCF modals provide a robust anharmonic basis, significantly preferable to a harmonic oscillator basis. The VSCF modals can themselves be obtained using a primitive basis through various methods. However, this is a straightforward numerical step that can be performed with high precision~\cite{toffoli_accurate_2011} and efficiency~\cite{hansen_new_2010} and will not be discussed further. A set of VSCF modals is also frequently used in calculating correlated vibrational wave functions such as \ac{fvci}, the exact solution for a given choice of basis,  and \ac{vcc} wave functions\cite{christiansen_vibrational_2004}, which will be used in some of our exploratory computations.

    Often no confusion is possible when discussing operators for a given mode in the following sections, and the mode indices will be omitted for conciseness.
        
    \subsection{Coordinates and kinetic energy operators}

    The choice of coordinates is important in vibrational computations, as these determine the functional form of the \ac{keo} and \ac{pes} which strongly affect the computational cost and adequacy of various approximations such as the usage of truncated $n$-body expansions of the \ac{pes}. 
    Rectilinear coordinates such as normal coordinates are a well-known, simple attractive choice in many contexts, and will be used in most computations here. 
    However, normal coordinates inherently have difficulties in describing large-amplitude motions of the nuclei, often leading to strong  inter-mode couplings in these cases. 
    For this reason, it can be advantageous to utilize curvilinear coordinates to describe the internal motion of molecules, 
    where angles and bond lengths naturally feature. 
     One such type of molecular motion is torsional motion that is essential both to describe some small molecules with hydrogen peroxide is the prototypical example as well as for understanding many biomolecular systems.
     Appropriate sets of curvilinear coordinates are generally expected to exhibit weaker couplings in the \ac{pes} in such cases.
    However, the vibrational \ac{keo} potentially takes on a much more complicated form when expressed in curvilinear coordinates, involving, depending on the details, up to full mode couplings and many terms. 
    Thus, the usage of an exact \ac{keo} could become very expensive in both conventional and quantum computations.
    However, just as the \ac{pes} is almost always represented in practice in some restricted form providing sufficient quality, so can the vibrational \ac{keo}. Thus, following general procedures
    providing numerically exact \acp{keo}~\cite{lauvergnat_exact_2002}, one can devise attractive approximated 
    \acp{keo} with limited mode-couplings and a much reduced number of terms in a sum of product form, similar to the \ac{pes} as discussed in next section.
   ~\cite{bader_efficient_2024}

    It should be mentioned in passing that for the exact \ac{keo} in normal coordinates, the so-called Watson operator~\cite{watson_simplification_1968} there are, in fact, also complicated coupling terms deriving from rovibrational coupling in addition to the simple $-\frac{1}{2}\frac{\partial^2}{\partial q^2}$ type of terms. These have somewhat similar complexities and solutions~\cite{toffoli_automatic_2007} as discussed above for curvilinear coordinates, but are fortunately less important, and are often neglected for larger systems, and will not be discussed further here.

    \subsection{The vibrational Hamiltonian}
    
    Within the Born--Oppenheimer approximation the molecular nuclei move according to a \ac{pes}. The \ac{pes} is from the outset unknown, but must be computed or represented in any computation addressing the nuclear potential.
    %but can be sampled through electronic structure theory calculations at specific nuclear geometries. In practice this can and has been done in many ways and it is beyond the scope of this work to review this. 
    We will limit ourselves to discussing one particular numerical realization, the \ac{sop} Hamiltonian. We choose to discuss and encode the \ac{sop} form because it provides a fairly general form to represent the vibrational Hamiltonian providing good chemical accuracy, efficiency, and flexibility. Furthermore, automatic procedures have been developed for obtaining SOP format Hamiltonians, including many of the \acp{pes} used in different contexts~\cite{sparta_adaptive_2009,ostrowski_tensor_2016,ostrowski_tensor_2016,klinting_employing_2018,schroder_transforming_2020,klinting_vibrational_2020}, and the numerical realization of the above-discussed \acp{keo}. We write the SOP Hamiltonian as 
    \begin{align}
        H & ={}  \sum_{\mathbf{m}}H^{\mathbf{m}} 
    \nonumber  \\
    & ={}  \sum_{\mathbf{m}} \sum_{\boldsymbol{o}^\mathbf{m}}c_{\boldsymbol{o}^\mathbf{m}}^\mathbf{m}\prod_{ \mathclap{m \in \mathbf{m}} } h^{m,o^m} 
    \nonumber  \\
        & = {} \sum_m \sum_{o^m} c^{m}_{o^m} h^{m,o^m}
        \nonumber  \\
        + & \sum_{m_1<m_2} \sum_{o^{m_1},o^{m_2}} 
        c^{m_1m_2}_{o^{m_1}o^{m_2}}
        h^{m_1,o^{m_1}}h^{m_2,o^{m_2}}
          \nonumber  \\
        + & \sum_{m_1<m_2<m_3} \sum_{o^{m_1},o^{m_2},o^{m_3}} 
        c^{m_1m_2m_3}_{o^{m_1}o^{m_2}o^{m_3}}
         \nonumber  \\
         & \times 
        h^{m_1,o^{m_1}}h^{m_2,o^{m_2}}h^{m_3,o^{m_3}}
         %  \nonumber  \\
        +  \cdots 
    \end{align}
    Here, the summation over $\mathbf{m}$ runs over mode combinations (MCs), where $\mathbf{m} = (m_1, m_2, \ldots)$ denotes the modes that are coupled in the following set of terms collected in $H^{\mathbf{m}}$. Note that the Hamiltonian includes one-mode terms with a single mode $(m)$ (no actual coupling), two-mode couplings between pairs of different modes $(m_1m_2)$, three-mode couplings with triples $(m_1m_2m_3)$, and so on. The highest level of MC is an important characteristic of a vibrational Hamiltonian, and we correspondingly refer to 2M and 3M Hamiltonians, etc., which are characterized by including only up to two-mode and three-mode terms, respectively.
    For each MC in $H$, there is a summation over a number of terms that, in the \ac{sop} format, can be written as summations for each mode in the MC. These summations are indexed by $o^m$ for mode $m$, referring to some set of operators $h^{m,o^m}$. For each MC, there is a set of coefficients $c_{\boldsymbol{o}^\mathbf{m}}^\mathbf{m}$, which give the weights of the different combinations of operators. Note that for each MC, $c_{\boldsymbol{o}^\mathbf{m}}^\mathbf{m}$ is generally a tensor. We will later discuss tensor decomposition as a means of optimizing the representation for our purposes.

    %$ \in [0, \dots{}, O^\mathbf{m} - 1]$.
    The present form includes a number of special cases. Taylor expansion \acp{pes} have $h^{m,o^m}$ operators that are simply of type $q^i$, with coordinate displacement $q$ to some power $i$. Taylor expansions may pose challenges owing to an uncertain or restricted radius of convergence and can become variationally unbound, thus potentially being less effective for precision tasks beyond a perturbative context and in quantum computation involving \ac{qpe}.
    For that reason, more general ways of obtaining the SOP Hamiltonian are attractive. 
    %Typically the \acp{pes} is obtained by sampling the individual MCs and after some manipulations fitting these a set of fit functions for each mode.
    Adaptive procedures such as the \ac{adga} have been developed to identify individual grids for each MC followed by fitting to selected arbitrary set of fitting basis functions~\cite{sparta_adaptive_2009,klinting_employing_2018} for each mode resulting in a \ac{pes} in \ac{sop} format.
    Typically physical-based adaptive approaches determine a PES that behaves physically sound within the domain it has been sampled, which in turn can be related to domain spanned by the vibrational basis. This leads to a \ac{pes} far
    beyond the quality of a single center Taylor expansion.~\cite{sparta_adaptive_2009,ostrowski_tensor_2016,ostrowski_tensor_2016,klinting_employing_2018,schroder_transforming_2020,klinting_vibrational_2020}

    A second quantization representation of the Hamiltonian is obtained by expressing the one-mode operators in terms of creations and annihilation operators\cite{christiansen_second_2004}
    \begin{align}
         h^{m,o^m} = \sum_{\mathclap{r^{m}s^{m}}} h_{r^{m} s^{m}}^{o^m} a_{r^m}\dag a_{s^m},
    \end{align}
    Here $h_{r^{m} s^{m}}^{o^m} = \langle r^m | h^{m,o^m}| s^m\rangle$ are one-mode integrals of the first quantization operators in one mode $h^{m,o^m}$ over the modals for mode $m$. 

    The above form of the \ac{sop} operator is useful for understanding its origin and connection to mode-couplings. The particular form is also useful for implementing wave function methods for conventional computers where it for example can be exploited that the basis of operators referred to by $o^m$ may be a small set of common operators for all MCs (most intuitive in the case of a common polynomial fit basis $q_m^i$), since then transformations with these can then be reused. In the present context it is, however, more desirable to have as simple mathematical form for the discussion and for the computations as few terms as possible.
    For the latter reason, we will in \cref{sec:tensor_decomposition} discuss tensor decomposition. This corresponds to transforming operators for each MC to a new set of optimal unique set of operators for each term.
    Furthermore, we may choose to renormalize the one-mode operators with the expansion coefficient to obtain a simplest possible form
    %
 %   HERE IS WHAT YOU HAD
 %   \begin{align} \label{eq:factorised_hamiltonian}
 %       H = \sum_{\mathbf{m}} \sum_{o^\mathbf{m}} \bigotimes_{ \mathclap{ m\in\mathbf{ m} }} H_{1\textsc{m}}(\mathbf{m}, o^\mathbf{m}, m).
 %   \end{align}
 %   HERE IS WHAT YOU ALSO COULD DO - but does the big cross nomenclature add anything you didnt 
        \begin{align} \label{eq:factorised_hamiltonian}
        H = \sum_{t}  \bigotimes_{ \mathclap{ m\in\mathbf{ m}_t }} H_{1\textsc{m}}^{m,t}.
    \end{align}
    The latter simply express that the final Hamiltonian is a sum of terms (indexed $t$) where in each term there is a product of specific one-mode operators for the specific modes active for that term, where the active modes is provided by the set $\mathbf{ m}_t$

    In the following we will discuss various aspects of encoding the operation and it will be decisive to have concise nomenclature. The action of one-mode operators are central for the discussion. When confusion is not possible we will  discuss them using a simpler form with all mode, term, and  operator indices suppressed
    \begin{align} \label{eq:one-mode_operator}
         H_{1\textsc{m}} = \sum_{rs} h_{rs} a_{r}\dag a_{s}.
    \end{align}
    where each sums run over $N_m$ modals for that mode.
    As Hamiltonian is hermitian and real, we will most often work with real hermitian operators in a real basis case, so $h_{rs} = h_{sr}$, and thus it is sometimes convenient to use the form 
    \begin{align}
         H_{1\textsc{m}} & = \frac{1}{2}\sum_{rs} h_{rs} (a_{r}\dag a_{s} + a_{s}\dag a_{r}) 
    \end{align}
    In other discussions it is the terms for specific mode combinations are in focus and we note we can also write
    \begin{align} \label{eq:sop_hamiltonian}
        H & ={}  \sum_{\mathbf{m}}H^{\mathbf{m}} =
        \sum_{\mathbf{m}} \sum_{ \mathclap{t^{\mathbf{m}}=1} }^{N_\textsc{t}^{\mathbf{m}}}
        H^{\mathbf{m}}_{t^{\mathbf{m}}}
    \end{align}
    emphasizing that for each MC $\mathbf{m}$ we have a sum over $N_\textsc{t}^{\mathbf{m}}$ terms. 
    
    \subsection{Qubitization}
    \label{sec:qpe}
    
    We encode the vibrational Hamiltonian $H = \sum_{i} c_{i} U_{i}$ represented as a \ac{lcu} where $U_{i}$ is a unitary operator and $c_i \in \R$. The \ac{lcu} is encoded using qubitization which encodes the Hamiltonian as a Szegedy walk operator
    \begin{align}
        W = R \times \mathcal{B}[H] = e^{ \pm i\arccos(H/\alpha)}
        \label{eq:qubitization}
    \end{align}
    where
    \begin{align} \label{eq:lcu_norm}
        \alpha = \sum_{i} |c_{i}|
    \end{align}
    is the sum of \ac{lcu} coefficients, which we will also refer to as the \ac{lcu} norm. We denote $\mathcal{C}[\mathcal{B}[H]]$ as the cost of block encoding the \ac{lcu} denoted by $\mathcal{B}[H]$ such that the query complexity of qubitization may be written as~\cite{low_hamiltonian_2019,babbush_encoding_2018}
    \begin{align}
        \mathcal{O} \left(\frac{\alpha}{\epsilon}\mathcal{C}[\mathcal{B}[H]] \right).
    \end{align}
    Since the overhead in performing T gates in standard fault-tolerant gate sets such as Clifford+T is much larger than that of Clifford gates~\cite{babbush_encoding_2018,bravyi_quantum_1998,fowler_bridge_2013,fowler_surface_2012}, we only consider the reduction of T gates in the encoding algorithms. Since the T gates also appear in the standard decomposition of the Toffoli gates in the quantum circuits used for block encodings, we count Toffoli gates in the cost functions of block encodings in this work.

    \section{Qubit representation of the bosonic operators} \label{sec:encoding}
    
    In order to obtain an \ac{lcu} representation for \cref{eq:sop_hamiltonian}, one must map the bosonic creation and annihilation operators into a qubit space representation. Several studies have investigated different encodings of vibrational degrees of freedom with a range of different conclusions depending on the type of operators to be encoded~\cite{ollitrault_hardware_2020,sawaya_resource-efficient_2020}. In this work, we utilize the direct encoding of vibrational degrees of freedom which yields a reasonable tradeoff between circuit and qubit complexity. With this choice a qubit is needed for each modal of each vibrational mode, i.e.
    \begin{align} \label{eq:vibrational_qubits}
        \mathcal{N}_\textup{vib} = \sum_{ m } N_m,
    \end{align}
    with the sum running over all modes.

    \subsection{Qubit representation of the creation and annihilation operators}
    Using the direct encoding of each modal, the bosonic operators transform such that~\cite{ollitrault_hardware_2020,sawaya_resource-efficient_2020,majland_optimizing_2023}
    \begin{align} \label{eq:bosonic_anni_encoding}
        \anni{}{r}, \crea{}{r} \rightarrow \sigma_{r}^{\mp}=\frac{1}{ 2}(\sigma_{r}^{x} \pm i\sigma_{r}^{y}).
    \end{align}
    where $\sigma_{r}^{-}$ ($\sigma_{r}^{+}$) is the Pauli annihilation (creation) operator and $\sigma_{r}^{x/y}$ are the Pauli $X/Y$ operators of the qubit handling the occupation of modal $r$.
    
    In electronic structure theory, the Majorana representation is typically used since it provides beneficial properties for encoding~\cite{von_burg_quantum_2021,steudtner_fault-tolerant_2023}. However, the bosonic Majorana representation corresponds to the direct encoding of the modals and thus the encodings are equal.
    
    Define the bosonic Majorana operators as
    \begin{subequations}
    \begin{align}
        \gamma_{r}^{0} &= \anni{}{r}+\crea{}{r}, \\
        \gamma_{r}^{1} &= -i(\anni{}{r}-\crea{}{r}).
    \end{align}   
    \end{subequations}
    which may be inverted such that
    \begin{align}
        \anni{}{r}=\frac{1}{2}(\gamma^{0}_{r}+i\gamma^{1}_{r}),
    \end{align}
    which simply corresponds to the Pauli matrices where $\sigma_{r}^{x} \leftrightarrow \gamma^{0}_{r}$ and $\sigma_{r}^{y} \leftrightarrow \gamma^{1}_{r}$.
    
     Using the transformations in \cref{eq:bosonic_anni_encoding}, the terms of the one-mode operators transform in \cref{eq:one-mode_operator} such that
    \begin{align}
    \begin{aligned}
        \crea{}{r}\anni{}{s} + \crea{}{s}\anni{}{r} &\rightarrow \sigma_{r}^{+}\sigma_{s} ^{-}+\sigma_{s}^{+}\sigma_{r} ^{-} \\
        &= \frac{1}{2} \left( \sigma_{r}^{x}\sigma_{s}^{x} + \sigma_{r}^{y}\sigma_{s}^{y} + \frac{i}{2}[\sigma_{r}^{x},\sigma_{s}^{y}] + \frac{i}{2}[\sigma_{s}^{x},\sigma_{r}^{y}] \right) \\
        &= \frac{1}{2} \big( \sigma_{r}^{x}\sigma_{s}^{x}+\sigma_{r}^{y}\sigma_{s}^{y} - 2 \sigma_{r}^{z}\delta_{rs} \big).
        \label{eq:one_mode_operator_expansions}
    \end{aligned}
    \end{align}

    As the Pauli operators are unitary and involutory, any product of Pauli operators is itself unitary.
    With the above transformation, the one-mode operators may therefore be written on the \ac{lcu} form such that
    \begin{align}
        H_{1\textsc{m}} \rightarrow \sum_{i} c_{i} U_{i}
    \end{align}
    where $U_{i}$ are products of Pauli operators, and therefore unitary, while $c_i$ are real expansion coefficients. The \ac{lcu} norm is evaluated straightforwardly by inserting the expansion coefficients into \cref{eq:lcu_norm}.
    
    Since the Toffoli and qubit costs for implementing an \ac{lcu} depend significantly on the representation, we consider three representations which have different coefficients (different expansion coefficients) and different costs of implementation for the unitary operators of the \ac{lcu}.
    
    We will use the notation $\mathcal{C}[\cdot]$ to denote the Toffoli gate cost of the operator in the brackets. Likewise, $\mathcal{N}[\cdot]$ denotes the number of qubits required to implement the operator in the brackets. We choose to partition the qubits into four sets; vibrational, readout, encoding and ancilla. $\mathcal{N}_\textup{vib}$ is the number of qubits required to store the vibrational state, which we refer to as the vibrational qubits, $\mathcal{N}_\textup{readout}$, is the size of the readout register, and $\mathcal{N}_\textup{enc}$ is the number of extra qubits required to block encode the \ac{lcu}.
    Several subcircuits require a certain number of ancillary qubits, the number of which is denoted $\mathcal{N}_\textup{anc}$. It is important to distinguish the encoding and ancilla registers as the reflection operator $R$ involves the encoding register, but not the ancilla register~\cite{babbush_encoding_2018,von_burg_quantum_2021}.

    \subsection{Quadratic representation}
    The quadratic, and the later defined triangular, representation is defined as an expansion of the bosonic operators into qubit operators using \cref{eq:one_mode_operator_expansions}. The advantage of this approach is the negligible cost of implementing the Pauli operators as they require zero Toffoli gates. Therefore, the only Toffoli cost arises from state preparation and the multiplex operation used to encode the sum of Pauli operators.

    Let $\sigma_{r}^{\alpha}$, where $\alpha\in\{x,y\}$, denote a $\sigma_{r}^{x}$ or $\sigma_{r}^{y}$ operator. Using \cref{eq:one_mode_operator_expansions}, the one-mode operators may be written as
    \begin{align} 
    \label{eq:one_mode_operator_quad}
    \begin{aligned}
        H_{1\textsc{m}}^\textsc{q} 
        &=
        \sum_{rs} h_{rs} \crea{}{r} \anni{}{s} \\
        &= \frac{1}{2} \sum_{rs} h_{rs} (\crea{}{r} \anni{}{s} + \crea{}{s} \anni{}{r}) \\
        %&=\frac{1}{4}\sum_{rs}h_{rs}(\sigma_{r}^{x}\sigma_{s}^{x}+\sigma_{r}^{y}\sigma_{s}^{y}+i[\sigma_{r}^{x},\sigma_{s}^{y}])\\
        &= \frac{1}{4} \sum_{rs}  h_{rs} ( \sigma_{r}^{x} \sigma_{s}^{x} + \sigma_{r}^{y} \sigma_{s}^{y} - 2 \sigma_{r}^{z} \delta_{rs} ) \\
        &= \frac{1}{4} \sum_{rs}  h_{rs} \sum_{ \mathclap{ \alpha\in\{x,y\} } } \sigma_{r}^{\alpha} \sigma_{s}^{\alpha} - \frac{1}{ 2} \sum_{r}  h_{rr} \sigma_{r}^{z}.
    \end{aligned}
    \end{align}
    The \ac{lcu} norm of the quadratic representation may be written as
    \begin{align}
        \alpha( H_{1\textsc{m}}^\textsc{q}) = \frac{1}{2} \sum_{rs} | h_{rs} | + \frac{1}{2} \sum_{r} | h_{rr} |.
    \end{align}
    The number of \ac{lcu} coefficients to be loaded is simply
    \begin{align} \label{eq:quadradratic_lcu_coeff_number}
        N_\textup{coef} [H_{1\textsc{m}}^\textsc{q}] = N_m^2.
    \end{align}

    \subsection{Triangular representation}
    The triangular representation is obtained from the quadratic representation by restricting the sum to the upper/lower triangular matrix elements. The upper triangular part is related to the lower by symmetry.
    
    The advantage of this approach, like the quadratic representation, is the negligible cost of implementing the Pauli operators, hence the only Toffoli cost arises from state preparation and the multiplex operations. In addition, state preparation is only required for the upper/lower triangular matrix elements with the triangular representation.

    As the one-mode operators are Hermitian, $\sigma_{r}^{\alpha} \sigma_{s}^{\alpha} = \sigma_{s}^{\alpha} \sigma_{r}^{\alpha}$ for $r \neq s$ and $\sigma_{r}^{\alpha} \sigma_{r}^{\alpha} = \id$ the $rs$ summation in \cref{eq:one_mode_operator_quad} may be restricted to only include triangular terms such that
    \begin{align} \label{eq:one_mode_operator_tri}
        H_{1\textsc{m}}^\textsc{t}
            &= \frac{1}{2} \sum_{r<s} h_{rs} \sum_{ \mathclap{ \alpha\in\{x,y\} } } \sigma_{r}^{\alpha} \sigma_{s}^{ \alpha} + \frac{1}{2} \sum_{r} h_{rr}  \left( \id - \sigma_{r}^{z} \right).
    \end{align}
    The \ac{lcu} norm of the triangular representation may be written as
    \begin{align}
        \alpha( H_{1\textsc{m}}^\textsc{t}) = \sum_{r \leq s} | h_{rs} | = \alpha( H_{1\textsc{m}}^\textsc{q}).
    \end{align}
    This implies that the number of \ac{lcu} coefficients to be loaded is
    \begin{align} \label{eq:triangular_lcu_coeff_number}
        N_\textup{coef} [H_{1\textsc{m}}^\textsc{t}] = N_m \frac{N_m + 1}{2}.
    \end{align}
    
    \subsection{Diagonal representation{\label{diagrep}}}
    The diagonal representation amounts to a diagonalization of the one-mode operators such that one-mode eigenvectors are encoded using basis transformed qubit operators. The one-mode operators may be diagonalized such that
    \begin{align}
    \label{eq:one_mode_operator_diag}
    \begin{aligned}
                 H_{1\textsc{m}}^\textsc{d} &= \frac{1}{4} \sum_j \sum_{rs} \lambda_{j} U_{rj} U_{js} \sum_{ \mathclap{ \alpha\in\{x,y\} }} \sigma_{r}^{\alpha} \sigma_{s}^{ \alpha} - \frac{1}{2} \sum_{r}  h_{rr}\sigma_{r}^{z} \\
                &= \frac{1}{4} \sum_j \sum_{ \mathclap{ \substack{ \alpha\in \\ \{x,y\} }}} \lambda_{j} \left( \sum_{r} c_{r}^{(j)} \sigma_{r}^{\alpha} \right) \left( \sum_{s} d_{s}^{(j)} \sigma_{s}^{ \alpha} \right) \\
                &- \frac{1}{2} \sum_{r}  h_{rr} \sigma_{r}^{z} \\
                &= \frac{1}{4} \sum_{j} \lambda_{j} \sum_{ \mathclap{ \alpha\in\{x,y\} }} \tau_{j}^{\alpha} \upsilon_{j}^{ \alpha} - \frac{1}{2} \sum_{r}  h_{rr} \sigma_{r}^{z}, \\
    \end{aligned}
    \end{align}
    where $c^{(j)}$ is the $j^\text{th}$ column and $d^{(j)}$ the  $j^\text{th}$ row of $U$, and
    \begin{subequations} \label{eq:pauli_transformed}
    \begin{align}
        \tau_{j}^{\alpha} &= \sum_{r} c_{r}^{(j)} \sigma_{r}^ {\alpha}, \\
        \upsilon_{j}^{\alpha} &= \sum_{s} d_{s}^{(j)} \sigma_{s}^{\alpha}.
    \end{align}
    \end{subequations}
    Thus, $\tau$ and $\upsilon$ are different transformations of the Pauli operators. Note that $\tau_{j}^{\alpha}$ and $\upsilon_{j}^{\alpha}$ depend on the particular one-mode operator, and herein the particular mode-combination and operator indices %$o^\mathbf{m}$ 
    through the basis transformations within them.
    
    The \ac{lcu} norm of the diagonal representation may be written as
    \begin{align}
        \alpha( H_{1\textsc{m}}^\textsc{d} ) = \frac{1}{2}\sum_{j} |\lambda_{j}| +\frac{1}{2} \sum_{r} | h_{rr}|.
    \end{align}     

    In comparison to the triangular representation, the unitary operators of the \ac{lcu} require Toffoli gates since basis transformed qubit operators require Toffoli gates. However, the number of coefficients for the state preparation scales with the number of eigenvalues of the one-mode operators as $\mathcal{O}(N)$ compared to $\mathcal{O}(N^{2})$ for the triangular representation, with $N$ being the dimension of the matrix.

    In addition to lowering the number of coefficients, Ref.~\cite{von_burg_quantum_2021} shows that the Schatten norm (sum of absolute values of eigenvalues) for a Hermitian matrix of dimension $N$ may be up to a factor of $N$ smaller than the entry-wise \ac{lcu} norm (quadratic/triangular representation). Thus, the \ac{lcu} norm of the diagonal representation may be up to a factor of $N$ smaller than the quadratic/triangular representation. However, the reduction in \ac{lcu} norm must be compared to the increase in gate costs for implementing the basis transformations.
    
    While the diagonal representation combined with low rank approximations of two-electron integrals in electronic structure~\cite{von_burg_quantum_2021,berry_qubitization_2019,steudtner_fault-tolerant_2023} was shown to provide significant reductions of \ac{lcu} norms, such \ac{lcu} norm reductions may not be readily obtained in the vibrational context for several reasons. In fact, for all systems considered in this work, the triangular representation turned out to require fewest gates.
    
    \paragraph{Size of one-mode operators}
    
    One-mode operator matrices are considerably smaller than the two-electron integrals, which scale with the number of electronic spin-orbitals, whereas the one-mode operator coefficients scale with the number of modals per mode, which does not scale with system size. In large-scale calculations with low temperatures, the number of modals per mode may be in the range of ten whereas the number of spin-orbitals may be hundreds scaling linearly with the size of the system. In the case of high temperatures and/or very anharmonic systems, however, it may be relevant with a larger number of modals per mode, say one hundred, and thus the \ac{lcu} norm reductions may be significant.
    
    \paragraph{Rank of one-mode operators}
    
    One-mode operators are not low-rank and thus truncating the terms in the one-mode operators may result in insufficient accuracy.

    \paragraph{Non-zero cost of basis transformations}
    
    The basis transformed qubit operators thirdly have non-zero Toffoli cost since these basis transformations must be encoded. Therefore, despite having fewer terms and a smaller \ac{lcu} norm, the diagonal representation could be inferior to the triangular representation depending on the \ac{lcu} norm reduction and the cost of implementing the basis transformed qubit operators.
   
    \section{Encoding of the sum-over-product operators}

    Block encoding of the Hamiltonian is obtained through a \aclp{sop} of block encodings for the individual one-mode operators. We thus use the Hamiltonian in Eq. (\ref{eq:factorised_hamiltonian})
    where the one mode operators are one of the various representations
    described above with the symbol $H_{1\textsc{m}} \in \{H_{1\textsc{m}}^\textsc{q}, H_{1\textsc{m}}^\textsc{t}, H_{1\textsc{m}}^\textsc{d}\}$. 
    \cref{sec:one-mode_operators} outlines the model used to compute Toffoli gates and qubit count for the one-mode operator block encodings. \cref{sec:one-mode_products} describes the implementation of the product of one-mode matrices and the implementation of the outer mode-coupling sum ($\mathbf{m}$) and the sum over expansion terms ($o^\mathbf{m}$) is described in \cref{sec:cp_rank_sum}. Additionally, \cref{sec:one-mode_products,sec:cp_rank_sum} outline the cost model of the full block encoding in terms of the one-mode cost functions.

    Note that when dealing with distinguishable, bosonic degrees of freedom, as in the case of vibrational theory, the operator Hilbert space factorises to a direct product space. Many sums add terms for each basis elements for a single degree of freedom, i.e. the size of the sums is typically the number of modals within a mode and not the total number of basis elements. The number of modals within each mode is modest, compared with the basis sizes common e.g. in electronic structure theory, typically on the order of unity to a few tenths, in which case it is typically insufficient to only consider the asymptotic scaling of circuits since linear and logarithmic terms may be important. However, in the following sections, we present only asymptotic scalings of the costs and discuss the more complicated exact cost functions in the Appendix.
     
    \subsection{One-mode operators} \label{sec:one-mode_operators}

    The block encoding for an operator $H_{1\textsc{m}}$ consists of \textsc{prepare} ($P$) and \textsc{select} ($S$) operators such that
    \begin{align}
        \mathcal{B} [H_{1\textsc{m}}] = P^{\dagger}SP
    \end{align}
    where
    \begin{align}
        \prescript{}{\textup{enc}}{\bra{0}} \mathcal{B}[H_{1\textsc{m}}] \ket{0}_\textup{enc} = \frac{H_{1\textsc{m}}}{\alpha},
    \end{align}
    and $\alpha$ is the \ac{lcu} norm of $H_{1\textsc{m}}$. That is, upon projection of the encoding qubits onto their zero state $\mathcal{B}[H_{1\textsc{m}}]$ effectively applies $H_{1\textsc{m}}/\alpha$ to the vibrational subspace.
    
    Thus, block encoding consists of two \textsc{prepare} operators and one \textsc{select}, hence the Toffoli gates model for the one-mode operators is
    \begin{align}
        \mathcal{C} \big[ \mathcal{B} [ H_{1\textsc{m}} ] \big] = \mathcal{C} [ P\dag ] + \mathcal{C} [ S ] + \mathcal{C} [ P ].
    \end{align}
    Note that using measurement-based uncomputation~\cite{berry_qubitization_2019} $\mathcal{C} [ P\dag ] \leq \mathcal{C} [ P ]$.

    The \textsc{prepare} operator for the vibrational problem may be implemented using the \textsc{subprepare} circuit described in Ref.~\cite{babbush_encoding_2018} and \textsc{select} can be implemented using unary iteration also described in Ref.~\cite{babbush_encoding_2018}.
    The Toffoli gates and qubit counts thus depend on the number of operator coefficients, that must be loaded from a classical database $N_\textup{coef}$ and the number of qubits ($\mu$) used to store the Hamiltonian coefficient ($h_{rs}$) floating point numbers.
    
    In Refs.~\cite{low_trading_2018,berry_qubitization_2019,von_burg_quantum_2021}, data look-up methods using the \textsc{selectswap} circuit are described that use additional qubits to decrease the Toffoli cost front factor of the leading-order (linear) term at the cost of introducing a term linear in the size of the output. To compute the data look-up, $\lambda_\textsc{c} - 1$ copies of the output register is required. The following uncomputation uses $\lambda_\textsc{u}$ ancilla qubits. Importantly, $\lambda_\textsc{c}$ is a number of registers, whereas $\lambda_\textsc{u}$ is a number of qubits and the two may assume different numeric values.

    However, with the relatively small modal subspaces typically used for vibrational wave functions compared to electronic wave functions, none of the benchmark computations performed in this work exhibited any benefit from utilizing the \textsc{selectswap}.
    
    It is possible to reuse the qubits allocated for \textsc{prepare} during \textsc{select} and the uncomputation of \textsc{prepare}. Additionally, it is shown in \cref{app:one-mode_operator} that the number of qubits needed for \textsc{prepare} is sufficient for the rest of the block encoding when using the quadratic/triangular representation, whereas the diagonal representation requires additional qubits in which to store the basis rotation angles.
    
    Detailed descriptions of the circuits and derivations of the cost functions can be seen in \cref{app:one-mode_operator}.

    \newcommand{\quadraticToffoliCost}{$\mathcal{O}( N_m^2 + \mu)$}
    \newcommand{\quadraticToffoliCostLambda}{$\mathcal{O} ( N_m^2 \left[1 + \frac{1}{\lambda} \right] + \mu\lambda )$}
    \newcommand{\quadraticQubitsLambda}{$\mathcal{O} ( \log_{2} (N_m) - \lg (\lambda) + \mu ( \lambda + 1) )$}
    \newcommand{\quadraticQubitsStandard}{$\mathcal{O}(\lg ( N_m )+\mu)$}

    \newcommand{\diagonalToffoliCost}{$\mathcal{O}(\beta N_m^2 + \mu)$}
    \newcommand{\diagonalToffoliCostLambda}{$\mathcal{O} (\beta N_m^2+ N_m \beta\lambda + \frac{N_m}{\lambda} + \mu\lambda )$}
    \newcommand{\diagonalQubitsStandard}{$\mathcal{O}(\lg ( N_m ) + \beta N_m + \mu  )$}
    \newcommand{\diagonalQubitsLambda}{$\mathcal{O}(\lg ( N_m ) - \lg (\lambda) + (\mu + \beta N_m) (\lambda + 1) )$}

    %\begin{center}
    
    \def \arraystretch{1.6} 
    \setlength{\tabcolsep}{.65em} 
    
    \begin{table*}[t]
    \centering 
    
    \begin{tabular} 
    {ccc} 
    \toprule 
    Representation & Toffoli gates (standard) & Toffoli gates (\textsc{selectswap}) \\ 
    \midrule 
    Quadratic/triangular & \quadraticToffoliCost & \quadraticToffoliCostLambda\\ 
    Diagonal & \diagonalToffoliCost & \diagonalToffoliCostLambda \\ 
    \bottomrule 
    \end{tabular}
    \caption{Asymptotic Toffoli costs for the quadratic, triangular and diagonal representations. $\mu$ represents the coefficient qubits, $\beta$ represents the basis transformation rotation angle qubits and $N_{m}$ is the number of modals per mode. The Toffoli gates (standard) represents the gate costs using the standard \textsc{select} circuit while the Toffoli gates (\textsc{selectswap}) represents the gate costs using the \textsc{selectswap} circuit where $\lambda$ is the parameter for the \textsc{selectswap} circuit.}
    \label{tab:asymptotic_costs_gates}
    \end{table*} 

    \begin{table*}[t]
    \centering 
    \begin{tabular} 
    {ccc} 
    \toprule 
    Representation & Qubits (standard) & Qubits (\textsc{selectswap}) \\ 
    \midrule 
    Quadratic/triangular & \quadraticQubitsStandard & \quadraticQubitsLambda \\ 
    Diagonal & \diagonalQubitsStandard & \diagonalQubitsLambda \\ 
    \bottomrule 
    \end{tabular} 
    
    \caption{Asymptotic qubits costs for the quadratic, triangular and diagonal representations analogous to \cref{tab:asymptotic_costs_gates}.
    }
    %$\mu$ is the coefficient qubits, $\beta$ is the basis transformation rotation angle qubits and $N_{m}$ is the amount of modals per mode. The encoding and coefficient qubits represent the qubits required to index the \ac{lcu} coefficients and the related binary representations. The ancillary qubits represent the qubits required for the implementation of the \textsc{selectswap} circuit.}
    \label{tab:asymptotic_costs_qubits} 
    \end{table*} 
    
    %\end{center}
    
    \subsubsection{Asymptotic costs for block encodings of one-mode operators}

    The asymptotic Toffoli cost of implementing the one mode operators in the quadratic and the triangular representations are presented in \cref{tab:asymptotic_costs_gates,tab:asymptotic_costs_qubits}. 
    
    The quadratic and triangular representations encode $\mathcal{O}(N_m^{2})$ coefficients which are stored in $\mathcal{O}(\mu)$ qubits and indexed through $\mathcal{O}(\lg(N_{m}))$ qubits yielding the Toffoli costs presented in \cref{tab:asymptotic_costs_gates,tab:asymptotic_costs_qubits} for both standard coherent alias sampling and $\textsc{selectswap}$ circuits.
    There $\lambda$ is the parameter for the $\textsc{selectswap}$ circuit.
    In the interest of simplicity, the scalings do not distinguish the different values $\lambda$ may take, e.g. for computation and uncomputation, as their assymptotic scalings are identical.

    The diagonal representation encodes $\mathcal{O}(N_m)$ coefficients stored in $\mathcal{O}(\mu)$ qubits and indexed through $\mathcal{O}(\lg(N_{m}))$ qubits. However, for each coefficient, the basis-transforming unitaries each require $N_m$-fold products of two-qubit rotations, discussed in detail in \cref{sec:basis_transformation_bosons}. The rotation angles require $\mathcal{O}(N_{m})$ data look ups using $\mathcal{O}(N_m \beta)$ qubits that must be uncomputed afterwards, as the qubits must be reused with a different set of rotation angles for the next one-mode operator.
    This results in a circuit whose leading order cost is $\mathcal{O}(\beta N_{m}^{2})$.
    
    Note that $\mu$ and $\beta$ are generally different as the operator coefficients and rotation angles do not necessarily require the same number of qubits to achieve the desired precision. This is discussed in \cref{app:sec:error_budget}.
    
    Both the linear and sublinear Toffoli costs for the respective representations are presented in \cref{tab:asymptotic_costs_gates,tab:asymptotic_costs_qubits}. Detailed derivations can be seen in \cref{app:cost_functions}.
        
    \subsection{Product of one-mode operators} \label{sec:one-mode_products}
    
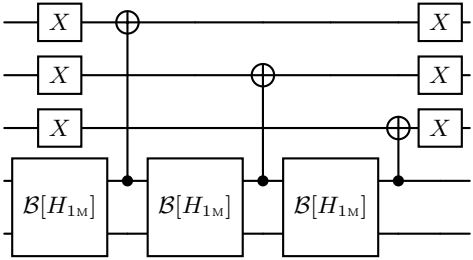
\begin{figure}
        \begin{center}
            \begin{quantikz}[align equals at=3,column sep=3pt, row sep={20pt,between origins}]
                & \gate{X} & \targ{} & & & & & \gate{X} & \\
                
                &\gate{X} &&& \targ{} &&& \gate{X}&\\
                &\gate{X} & & & & & \targ{}  & \gate{X} & \\
                & \gate[2]{\mathcal{B}[H_{1\textsc{m}}]} & \ctrl{-3}& \gate[2]{\mathcal{B}[H_{1\textsc{m}}]} & \ctrl{-2} & \gate[2]{\mathcal{B}[H_{1\textsc{m}}]}  & \ctrl{-1} & &\\
                & & & & & & & &\\ 
            \end{quantikz}
        \end{center}
        \caption{Quantum circuit for a three-way product of one-mode block encodings $H_{1\textsc{m}}$. The circuit can be adapted to any mode-coupling level, hence this type of circuit block encodes $H^{\mathbf{m}}_{t^{\mathbf{m}}} = \bigotimes_m H_{1\textsc{m}}(m)$ with additive cost $\mathcal{O}(N^2)$.}
        \label{fig:product}
    \end{figure}

    Products of one-mode operators may be implemented as products of block encodings, i.e. performing Kronecker products between the \textsc{select} and \textsc{prepare} operators for each one-mode operators in the product. However, such an approach will have a multiplicative cost $\mathcal{O}(N^{2n})$ where $n$ is the number of terms in the product. To mitigate the multiplicative scaling, one may utilize the approach in Refs.~\cite{von_burg_quantum_2021,cortes_fault-tolerant_2023}, using the circuit in \cref{fig:product}. The advantage of this approach is the additive cost of encoding products of block encodings, i.e. $\mathcal{O}(N^{2})$. 
    
    Let $|\mathbf{m}|$ denote the mode-coupling level of mode coupling $\mathbf{m}$, i.e. $n$ for an $n$-mode coupling. 
    %Using the notation $\mathcal{C}_{1\textsc{m}}(m) = $ and $\mathcal{N}_\textup{enc}^{1\textsc{m}}(m) = \mathcal{N}_\textup{enc} \big[ \mathcal{B}[H_{1\textsc{m}}(m)] \big]$, t
    The Toffoli cost of the block encoding of products is
    \begin{align}
        \mathcal{C} \big[ \mathcal{B}[ H^{\mathbf{m}}_{t^{\mathbf{m}}} ] \big] &= \sum_{ \mathclap{ m\in\mathbf{m} } } \mathcal{C} \big[ \mathcal{B}[H_{1\textsc{m}}] \big](m) + |\mathbf{m}|,
    \end{align}
    and the encoding and ancillary qubit cost is
    \begin{align}
        \mathcal{N}_\textup{enc} \big[ \mathcal{B}[ H^{\mathbf{m}}_{t^{\mathbf{m}}} ] \big] = \max_{m \in \mathbf{m}} \Big( \mathcal{N}_\textup{enc} \big[ \mathcal{B}[H_{1\textsc{m}}] \big] (m) \Big) + |\mathbf{m}|
    \end{align}
    and
    \begin{align}
        \mathcal{N}_\textup{anc} \big[ \mathcal{B}[ H^{\mathbf{m}}_{t^{\mathbf{m}}} ] \big] &= \max_{m \in \mathbf{m}} \Big( \mathcal{N}_\textup{anc} \big[ \mathcal{B}[H_{1\textsc{m}}] \big] (m) \Big).
    \end{align}
    %
    %Here $i \in \{\textup{enc}, \textup{anc} \}$.
    That is, the product requires an encoding qubit for each mode in the mode-coupling in addition to the largest number of qubits required by any of the one-mode operators.

    \subsection{Serial sum of block encodings} \label{sec:cp_rank_sum}
 
    The outer mode-coupling and operator term sums may be implemented as a linear combination of block encodings. Using an index register of size $\max_{\mathbf{m}} \lg( \ceil{|G|N_\textsc{t}^{\mathbf{m}}} )$, where $|G|$ denotes the number of mode couplings in the Hamiltonian, yields a total number of encoding qubits
    \begin{align}
    \begin{aligned}
        \mathcal{N}_\textup{enc} \big[ \mathcal{B}[ H ] \big] &= \max_{ \mathclap{\mathbf{m}, t^\mathbf{m}} } \Big( \mathcal{N}_\textup{enc} \big[ \mathcal{B}[H_{1\textsc{m}}] \big] (m) \Big)  \\
        &+ \max_{\mathbf{m}} \lg( \ceil{|G|N_\textsc{t}^{\mathbf{m}}} ).
    \end{aligned}
    \end{align}
    The notation $|G|$ is used, as it will later become the number of groups.
    The ancilla cost is the maximum number of ancillae in use at any one time, i.e.
    \begin{align}
        \mathcal{N}_\textup{anc} \big[ \mathcal{B}[ H ] \big] = \max_{ \mathclap{\mathbf{m}, t^\mathbf{m}} } \Big( \mathcal{N}_\textup{anc} \big[ \mathcal{B}[H_{1\textsc{m}}] \big] (m) \Big) .
    \end{align}

    The Toffoli cost of the sum is
    \begin{align} \label{eq:toffoli_cost}
    \begin{aligned}
        \mathcal{C} \big[ \mathcal{B}[H] \big]
        &= \sum_{\mathbf{m}} \sum_{t^\mathbf{m}} \mathcal{C} \big[ \mathcal{B}[ H^{\mathbf{m}}_{t^{\mathbf{m}}} ] \big].
    \end{aligned}
    \end{align}
    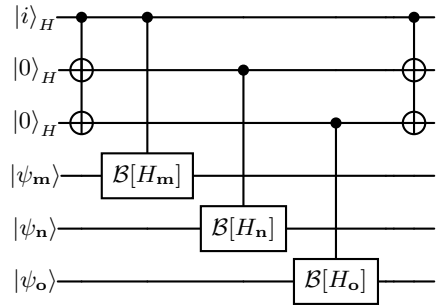
\begin{figure}
        \begin{center}
            \begin{quantikz}[align equals at=3,column sep=3pt, row sep={20pt,between origins}]
            % {30pt,between origins}]
                \ket{i}_{H} & \ctrl{2} & \ctrl{3} &&&&&\ctrl{2}& \\
                
                \ket{0}_{H} & \targ{} && \ctrl{3} &&&&\targ{}&\\

                \ket{0}_{H} & \targ{} &&&\ctrl{3}&&& \targ{}&\\

                \ket{\psi_{\mathbf{m}}} && \gate[1]{\mathcal{B}[H_{\mathbf{m}}]} &&&&&&\\

                \ket{\psi_{\mathbf{n}}} &&& \gate[1]{\mathcal{B}[H_{\mathbf{n}}]}  &&&&&\\ 
                
                \ket{\psi_{\mathbf{o}}} &&&& \gate[1]{\mathcal{B}[H_{\mathbf{o}}]} &&&&\\
                
            \end{quantikz}
        \end{center}
        \caption{Quantum circuit for the parallelized sum of block encodings for commuting operators $H_{G_{j}}=\sum_{\mathbf{m}\in G_{j}}H_{\mathbf{m}}$. The mode subspaces are represented by $\ket{\psi_\mathbf{m}}$.}
        \label{fig:parallelized_block_encodings}
    \end{figure}

    \subsection{Circuit parallelisation}
    A key feature of the vibrational problem is that one is concerned with distinguishable, bosonic degrees of freedom, hence the total Hilbert space factorises into a direct product space of one-mode Hilbert spaces. This implies that any operator $H_m$ in mode $m$ commutes with any operator $H_{m'}$ acting on another mode $m' \neq m$. This enables parallelisation of the block encodings across sets of non-overlapping modes.
    
    Let $\mathcal{G} = \{G_j\}$ denote a set of disjoint groups $G_j$ such that $\mathcal{G}$ spans the set of all mode-couplings, i.e. all terms in the Hamiltonian. Using these groups the Hamiltonian in \cref{eq:factorised_hamiltonian} may be written as
    \begin{align} \label{eq:parallelised_hamiltonian}
        H^\textup{par} &= \sum_{ G_j \in \mathcal{G} } \sum_{\mathbf{m} \in G_j} \sum_{t^\mathbf{m}} H^{\mathbf{m}}_{t^{\mathbf{m}}} = \sum_{ \mathclap{G_j \in \mathcal{G}} } H_{G_j}
    \end{align}
    where $H_{G_j}=\sum_{\mathbf{m} \in G_j} \sum_{t^\mathbf{m}} H^{\mathbf{m}}_{t^{\mathbf{m}}}$ is a sum of disjoint operators. The parallelized block encoding implements an additional outer summation compared to \cref{eq:factorised_hamiltonian} which encodes the summation of disjoint groups in addition to the summation of the mode coupling terms. The quantum circuit which implements the parallel summation of disjoint terms is presented in \cref{fig:parallelized_block_encodings}, and the details are described in \cref{sec:parallelization_block_encodings}.
        
    Since the disjoint groups are encoded in parallel, the circuit depth scales with the maximum cost operator in the group. Thus, the cost of the parallelized block encoding may be written as
    \begin{align} \label{eq:toffoli_cost_parallel}
        \mathcal{C} \big[ \mathcal{B} [ H^\textup{par} ] \big] = \sum_{ G_j \in \mathcal{G} } \max_{\mathbf{m} \in G_j} \mathcal{C} \big[ \mathcal{B}[ H_{\mathbf{m}} ] \big].
    \end{align}
    %
    %where $H_{\mathbf{m}} = \sum_{t^\mathbf{m}} H^{\mathbf{m}}_{t^{\mathbf{m}}}$. 
    The required number of qubits for the parallelized block encoding depends on the maximum required number of qubits for any group. In addition, the parallelized block encoding requires additional encoding qubits for performing a \textsc{fanout} operation. Thus, the required number of qubits may be written as
    \begin{align}
    \begin{aligned}
        \mathcal{N}_\textup{enc} \big[ \mathcal{B}[ H^\textup{par} ] \big] &= \max_{\mathclap{G_j \in \mathcal{G}}} \mathcal{N}_\textup{enc} \big[ \mathcal{B}[ H_{G_j}] \big] \\
        &+ \max_{\mathbf{m}} \lg( \ceil{|G|N_\textsc{t}^{\mathbf{m}} }) \\
        &\times \left( \max_{\mathbf{m}} \lg( \ceil{|G|N_\textsc{t}^{\mathbf{m}}} ) - 1 \right).
    \end{aligned}
    \end{align}
    All elements of a group requires it own ancillary register, hence
    \begin{align}
        \mathcal{N}_\textup{anc} \big[ \mathcal{B}[ H^\textup{par} ] \big] &= \max_{\mathclap{G_j \in \mathcal{G}}} \mathcal{N}_\textup{anc} \big[ \mathcal{B}[ H_{G_j}] \big].
    \end{align}

    \subsection{Grouping as graph coloring}
    \label{sec:grouping}

    The number of disjoint groups impacts the overall circuit depth as discussed above.
    Thus, reducing the total number of disjoint groups is essential for improving circuit depth.
    Interestingly, we observe that the grouping of disjoint operators can be reduced to the classical graph coloring~\cite{jensen_graph_1994} problem.
    Let us define a graph $G$, where nodes are labeled with operators and nodes of non-disjoint operators are connected with edges.
    A $k$-coloring of such a graph essentially partitions operators into $k$ disjoint groups.
    In other words, operators with the same color can be scheduled in the same layer i.e., in parallel.
    Computing the minimum number of partitions corresponds to finding the so-called chromatic number of the graph $G$.
    
    \paragraph{Algorithms}
    First, we establish a baseline by computing the existing disjoint groups in the given operator list.
    We propose a \emph{naive} algorithm, that iterates through operators in sequence and groups them if they are disjoint.
    If an operator is not disjoint with the current group, it is added to a new group and the iteration continues.
    While this algorithm takes linear time in the operator count, disjoint group count is close to the operator count.
    Thus, we still need graph coloring algorithms for better groupings.

    We investigate two such algorithms. First is an exact algorithm based on Classical Planning~\cite{ghallab_automated_2004}.
    Since graph coloring is NP-hard, exact algorithms are not scalable but they allow us to establish the optimal disjoint group counts.
    We encode graph coloring problem in Planning Domain Description Language (PDDL)~\cite{mcdermott_1998_2000}, a standard specification for classical planning.
    We can then use the state-of-the-art planners for solving our coloring problem.
    Essentially, the planner chooses a set of disjoint operators for each color.
    Thus, the effective plan length corresponds to the total number of colors needed.
    Using an optimal planner guarantees the optimal plan length thereby ensuring optimal graph coloring.
    In our experiments, we use SAT-based solver Madagascar~\cite{rintanen_madagascar_2014} with parallel plans for scalability.
    Second is an existing heuristic approach, we use the \emph{largest-first} greedy algorithm~\cite{welsh_upper_1967} available in NetworkX~\cite{hagberg_exploring_2008}.
    Greedy algorithm scales well with linear time complexity.
    Interestingly, we observed group counts close to optimal despite being a heuristic approach.

    \paragraph{Cost prioritization}

    Decomposing different operators results in different circuit depths.
    However optimizing for different costs of operators results in a harder numerical optimization problem.
    First, we again establish a baseline by considering all operators of same cost or zero cost.
    Thus, we can use graph coloring algorithms for partitioning without any adaptation.
    While straightforward, assuming operators are of same cost is an over approximation.
    For the second strategy, we apply a divide and conquer approach based on the operators costs.
    Instead of global graph coloring, we apply graph coloring for each operator set with the same cost.
    We start with the highest cost operators and proceed in the descending order based on cost.
    As we see in the experiments, the weighted approach improves the overall depth in the greedy algorithm.
    
    \section{Tensor decomposition of the vibrational Hamiltonian} \label{sec:tensor_decomposition}

    In order to reduce the number of terms in the vibrational Hamiltonian given by \cref{eq:sop_hamiltonian} and thus reduce the circuit depth of block encoding, low-rank approximations of the coupling tensors $c_{o^\mathbf{m}}^\mathbf{m}$ are utilized.
    While tensor decompositions for \ac{pes} representations have been pursued before~\cite{ostrowski_tensor_2016,schroder_transforming_2020}, our context and approach are different and will be briefly described in the following. For a general review on tensor decompositions used, we refer to~\cite{kolda_tensor_2009}.

    A reduction of terms for the one-mode part is trivially obtained 
    %to carry out through obtaining a single operator per mode 
    by summation of one-mode operators for the pure one-mode part for each mode.
    
    Two-mode couplings are represented as summations over two operators for mode pairs with coefficients being matrices, $c^{m_1m_2}_{o^{m_1}o^{m_2}}$, see the two-mode term of Eq.(\ref{eq:sop_hamiltonian}). Here, summations can be truncated using matrix \acp{svd} for the coefficient matrix $\mathbf{c}^{m_1m_2}$ for each pair of modes.
    
    The coupling coefficients for higher order mode-couplings are represented by higher order tensors, and thus tensor decomposition methods can be used to perform low-rank approximations.
    
    In this work, we utilize \ac{cp} decompositions using the \ac{cpals} method
    in order to obtain low-rank decompositions of higher-order mode-couplings through truncation of the \ac{cp} ranks. However, optimizing the \ac{cp} decomposition of large tensors can be numerically tricky and expensive. Thus, one may optionally perform an initial Tucker decomposition of the tensor to improve the stability and performance of \ac{cpals}. In this work, the Tucker decomposition uses the \ac{hooi} method~\cite{de_lathauwer_multilinear_2000,kolda_tensor_2009}. Since the dimension of the Tucker decomposition may be significantly lower than the original tensor, and subject to rigorous error bounds, stability, computational cost and performance of \ac{cpals} may be improved.
    Of course, standard \ac{hooi} relies on \acp{svd} of tensor unfoldings, which can become computationally demanding for very high-dimensional tensors. However, it is important to note that couplings beyond four modes are rarely encountered in practice, as computing such high-order terms in the \ac{pes} is both expensive and typically unjustified by their limited significance — particularly when a well-chosen coordinate system is used.
    
    More details on how the tensor decomposition techniques are used can be found in \cref{sec:tensor_decomposition_appendix}.
    
    We denote Tucker thresholds as $\epsilon_\textsc{t}$ and low-rank thresholds (\ac{cp} thresholds) as $\epsilon_\textsc{lr}$. We apply these to the standard Frobenius norm (denoted $\norm{\cdot}_2$) of the difference between the original and the decomposed tensor.  %For two-mode couplings the same $\epsilon_\textsc{t}$ is used.
    That is, the maximum error allowed for each tensor is $\epsilon_\textsc{lr}$.
    
    To relate the accuracy of the Hamiltonian after decomposition of all tensors to  accuracy of the Hamiltonian eigenvalues, we calculate the sum of tensor errors for different threshold values ($\epsilon_\textsc{lr}$). That is, with $c_{\mathbf{o}^\mathbf{m}}^{\mathbf{m}}$ and $\tilde{c}_{\mathbf{o}^\mathbf{m}}^{\mathbf{m}}$ denoting the elements of the original and decomposed tensors respectively, we use the error measure
    \begin{align} \label{eq:tensor_error}
	\epsilon_\textup{tensor} = \sum_{\mathbf{m}} \sum_{\mathbf{o}^\mathbf{m}} \norm{ c_{o^\mathbf{m}}^\mathbf{m} - \tilde{c}_{o^\mathbf{m}}^\mathbf{m} }_{2}.
    \end{align}

    In the worst case, $\epsilon_\textup{tensor}$ scales linearly with the number of terms in the summation and thus linear with respect to system size.
    However, \cref{sec:tensor_results} shows that such catastrophic error compounding is not observed in practice and the total tensor error is effectively controlled by the decomposition threshold $\epsilon_\textsc{lr}$.

    \section{Results}

    \subsection{\acs{qpe} costs for benchmark molecules}
    
    \def \arraystretch{1.4} 
    \setlength{\tabcolsep}{.45em} 
    \begin{table} 
    \centering 
    \begin{tabular} 
    {ccccc} 
    \toprule 
    \multirow{2}{*}{Molecule} & Modes/ & Toffoli & \multirow{2}{*}{Qubits} \\ 
    & Modals & gates & \\ 
    \midrule 
    Hydrogen peroxide (3M) & 6/5 & $\num{2.94e+12}$ & 430 \\ 
    Formaldehyde (2M) & 6/5 & $\num{2.00e+10}$ & 416 \\ 
    Formaldehyde (3M) & 6/5 & $\num{5.05e+11}$ & 442 \\ 
    Benzene (2M) & 30/3 & $\num{9.44e+09}$ & 1765 \\ 
    Benzene (3M) & 30/5 & $\num{2.73e+13}$ & 2285 \\ 
    Naphthalene (2M) & 48/3 & $\num{2.17e+10}$ & 2859 \\ 
    Naphthalene (3M) & 48/5 & $\num{5.27e+14}$ & 3975 \\ 
    PAH8 (2M) & 156/3 & $\num{2.01e+11}$ & 10191 \\ 
    \bottomrule 
    \end{tabular}
    \caption{QPE costs for formaldehyde, hydrogen peroxide, benzene, naphtalene and PAH8 using chemically accurate decomposed vibrational Hamiltonians using three modals per mode for all molecules. The decomposed vibrational Hamiltonians are truncated to ensure chemical accuracy and parallelization and grouping methods are utilized to reduce circuit depths.}
    \label{tab:results} 
    \end{table} 
    
    In \cref{tab:results}, we present the \ac{qpe} costs for formaldehyde, hydrogen peroxide, benzene (PAH1), naphthalene (PAH2) and PAH8 using both 2M and 3M \acp{pes} using three modals per mode for all molecules. The \ac{qpe} costs for all benchmark molecules are presented in \cref{tab:appendix_qpe_costs}, and the \ac{pah} molecular test series is defined in \cref{sec:compdetails}. The costs are calculated for chemically accurate vibrational Hamiltonians defined as the decomposed Hamiltonians with $\epsilon_\textsc{lr} = \SI{1e-8}{\hartree}$ for which the tensor errors were found to ensure chemical accuracy defined as $\epsilon = \SI{4.5e-6}{\hartree} = \SI{1.0}{\per\centi\metre}$. The \ac{cp} decompositions are calculated for Tucker decompositions with $\epsilon_\textsc{t}=10^{-10}$. The quantum circuits encoding the Hamiltonians utilize greedy encoding with weighted cost functions which were found to scale most efficiently for most systems.
    All molecules except for hydrogen peroxide are represented in normal coordinates where hydrogen peroxide is represented in polyspherical coordinates with decomposed kinetic energy operator. All molecules were constructed with high-level \acp{pes}. Details seen in \cref{tab:appendix_qpe_costs} and discussed in \cref{sec:compdetails}.

    In electronic structure theory, many papers attempt to estimate the required runtime of quantum circuits through compiling the logical gates to physical gates while assuming a specific type of quantum error correction code and properties of fault-tolerant quantum hardware~\cite{babbush_encoding_2018,lee_even_2021,motta_low_2021,von_burg_quantum_2021}. However, these considerations are primarily concerned with estimating potential quantum advantages by considering systems which are chemically relevant but classically intractable.
    
    In this work, we consider benchmark molecules that are suitable for investigating the performance of the algorithms, and not necessarily good candidates for potential quantum advantages for a chemically relevant vibrational systems. Depending on which quantum error correction model is assumed, however, the Toffoli gate runtime varies accordingly. With the quantum error correction model in Ref.~\cite{lee_even_2021}, the Toffoli gate runtime is estimated to be \SI{40}{\milli\second} such that any gate depth above $10^{10}$ Toffoli gates would require days to compute. This implies that all benchmark molecules utilized in this study would require several days. While these systems are not optimized chemically, for example in terms of coordinate systems and modal subspace dimensions for each mode,
    %and primitive basis sizes in the generation of the \ac{pes}, we claim that by optimizing such details
    the gate costs in \cref{tab:results} could be brought further down.

    \subsection{Mode-coupling level}

    \begin{figure*}[ht]
        \centering
        \includegraphics[width=\textwidth]{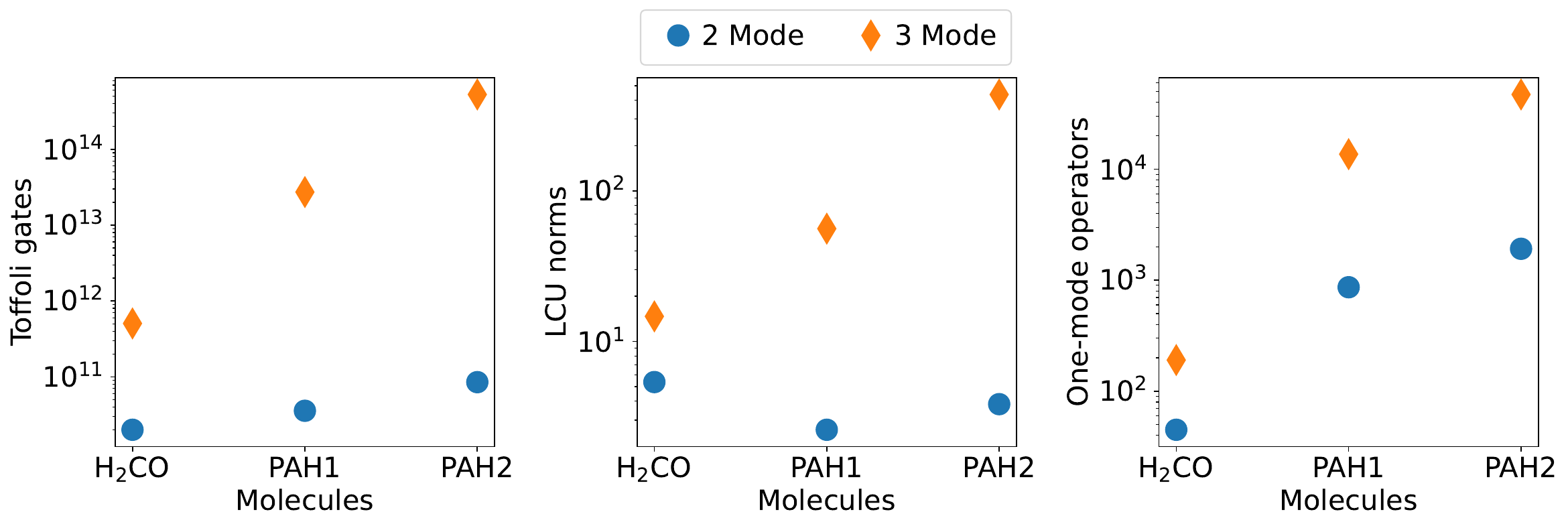}
        \caption{Costs for calculating the ground state energy for a series of \acp{pah} molecules with maximum \ac{cp} threshold yielding tensor errors ensuring at least chemical accuracy.}
        \label{fig:potential_energy_surfaces}
    \end{figure*}

    The maximum mode coupling of the \ac{pes} significantly impacts the number of terms and \ac{lcu} norm of the vibrational Hamiltonian. In order to investigate these dependencies, we calculate the gate costs, the number of terms and the \ac{lcu} norms for formaldehyde, benzene and naphtalene for both 2M and 3M \ac{pes} in \cref{fig:potential_energy_surfaces}. As can be seen in left figure of \cref{fig:potential_energy_surfaces}, the gate cost for the same molecule with 2M and 3M \ac{pes} varies significantly with more than three orders of magnitude in the case of naphtalene. Such rapid increases in gate costs would be expected since the number of terms increase strongly as a function of maximum mode-coupling level. In addition, such rapid increases with maximum mode coupling are also reflected in the center and right figures of \cref{fig:potential_energy_surfaces} for which both the \ac{lcu} norm and the number of one-mode operators in the vibrational Hamiltonian increase up to an order of magnitude for naphtalene. Somewhat surprisingly, the \ac{lcu} norm for formaldehyde is larger than both that of benzene and naphtalene in the case of 2M \ac{pes}. In addition, the \ac{lcu} norm variation between 2M and 3M \ac{pes} is much smaller than that for benzene and naphtalene.

    \subsection{Parallelization and grouping algorithms}
    
    \begin{figure}[ht]
        \centering
        \includegraphics[width=\columnwidth]{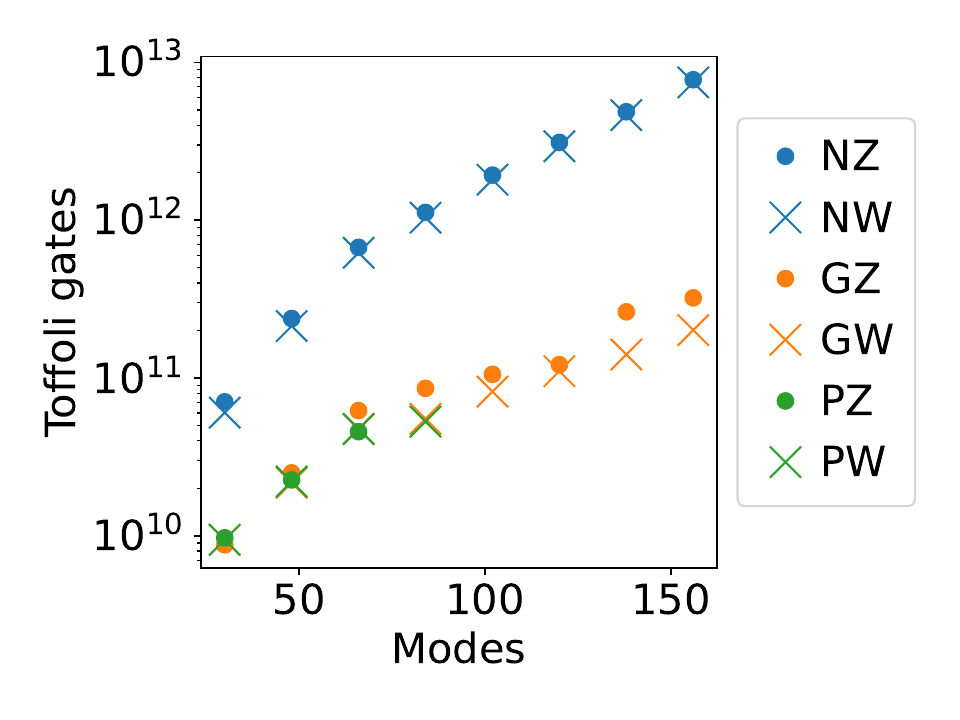}
        \caption{Costs for calculating the ground state energy for a series of \acp{pah} molecules with maximum \ac{cp} threshold yielding tensor errors ensuring at least chemical accuracy. The labels reflect the encoding and cost function such that $N=\textrm{naive}, G=\textrm{greedy}, P=\textrm{planning}$ and $Z=\textrm{zero}, W=\textrm{weighted}$.}
        \label{fig:mode_scaling}
    \end{figure}
    
    To investigate the impact of parallelization and grouping as a function of system size, we calculate \ac{qpe} costs for a series of \ac{pah} molecules with mode couplings $n=2$ as presented in \cref{fig:mode_scaling}. The cost computations are carried out for three different encodings, the naive, greedy and planning, with two different cost functions, zero and weighted. As can be seen, the greedy and planning algorithms outperform the naive algorithm by orders of magnitude with increasing performance as a function of system size. Since the planning algorithm scales exponentially, it does not scale to larger than PAH4 while the greedy algorithm scales efficiently to more than a hundred modes. While the encodings significantly impact the gate counts, the cost functions do not exhibit significant performance differences for the naive and planning encodings while some performance differences are exhibited for the greedy algorithm.

    \subsection{Curvilinear coordinates and one-mode operator representations}
    
    \begin{figure}[ht]
        \centering
        \includegraphics[width=\columnwidth]{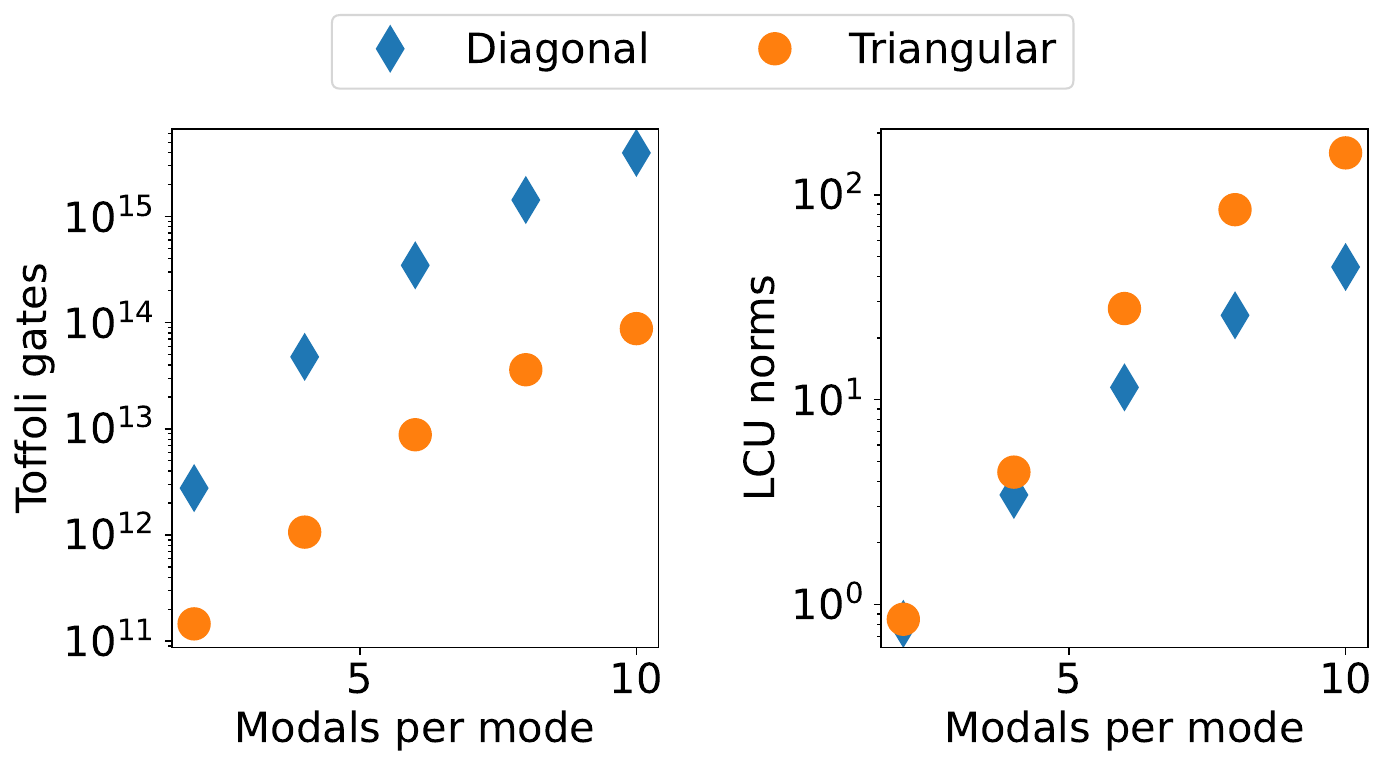}
        \caption{Left figure: \ac{qpe} costs for hydrogen peroxide in polyspherical coordinates as a function of the number of modals per mode for both the triangular and diagonal representations. Right figure: \ac{lcu} norm for the vibrational Hamiltonians in both the triangular and diagonal representations.}
        \label{fig:modal_scaling}
    \end{figure}

    For all other computations in this work, we utilize rectilinear coordinates for simplicity in the construction of the \ac{pes}. However, since the choice of coordinates is important in vibrational computations, we investigate the impact of utilizing curvilinear coordinates considering the hydrogen peroxide in polyspherical coordinates. 
   
    While the \ac{keo} is full-mode coupled analytically, we use a \acp{keo} with limited mode-couplings and a reduced number of terms as described in Ref.~\cite{bader_efficient_2024}. Specifically, the \ac{keo} is approximated using a low-order \( n \)-mode expansion of the so-called \( G \)-matrix. This approach yields reliable and systematically improvable approximations of the full-mode \ac{keo}.
    In Ref.~\cite{bader_efficient_2024}, several options were explored and we employ a 1-mode expansion \( G \), resulting in an overall 3-body kinetic operator which was shown to give sufficient accuracy. Given that the torsional mode is significantly more anharmonic than any other mode considered in this study, it is pertinent to use a larger modal basis for such a system. Consequently, we investigate the impact of increasing the number of modals for hydrogen peroxide. For simplicity, we use the same number of modals for all modes, although this consideration is most crucial for the dihedral angle. In addition, we investigate the performance of both the triangular and diagonal representations as a function of modals. 
    
    We calculate the \ac{qpe} costs for hydrogen peroxide with modals ranging from 2 to 10 modals per mode. The results are presented in \cref{fig:modal_scaling}. As can be seen, the gate counts increase rapidly as a function of increasing modal subspaces, which similarly increases for the \ac{lcu} norms. As would be expected, the differences in \ac{lcu} norm between the triangular and diagonal representations differ increasingly as a function of modals per mode. However, the difference in gate counts for the encodings is considerable, and the triangular representation appears to yield the lowest gate count for all modals per mode. Thus, despite the reductions in one norm provided by the diagonal representation, the cost of basis transformations in the diagonal representation appears to be significantly larger than the reductions in \ac{lcu} norms.

    \subsection{Tensor decomposition} \label{sec:tensor_results}
    
    To investigate the impact of tensor decomposition, 
    we study three metrics; the tensor error defined by \cref{eq:tensor_error}, the energy or eigenvalue error and the QPE cost metrics.

    Specifically, we calculate the \ac{qpe} costs for formaldehyde and PAH2 (naphthalene) with \acp{pes} at the three-mode-coupled level at different low-rank thresholds using five modals per mode for both molecules.
    Formaldehyde is sufficiently small such that \ac{fvci} calculations can be used to estimate the lowest eigenvalue of the Hamiltonian, hence the eigenvalue error can be computed without approximations.
    As the effectiveness of tensor decomposition supposedly increases with system size,
    PAH2 is used to test this statement.
    In this case, energy errors are computed at the VCC[2pt3] level of theory,
    meaning VCC with one-, two- and three-mode couplings where the VCC three-mode equations, solved iteratively, are approximated inspired by perturbation theory~\cite{seidler_vibrational_2009}.
    
    %, we perform high-level correlated calculations for the decomposed vibrational Hamiltonian. The energy obtained from the decomposed Hamiltonian is compared to that of the exact vibrational Hamiltonian.
    We define the energy error of the approximate vibrational Hamiltonians as
    \begin{align}
        \Delta E(\epsilon_\textsc{t},\epsilon_\textsc{lr})=|E_\textup{exact}-E(\epsilon_\textsc{t},\epsilon_\textsc{lr})|
    \end{align}
    where $E_\textup{exact}$ is the energy of the exact vibrational Hamiltonian and $E(\epsilon_\textsc{t},\epsilon_\textsc{lr})$ is the energy of the approximate vibrational Hamiltonian.

    \begin{figure}
        \centering
        \includegraphics[width=\columnwidth]{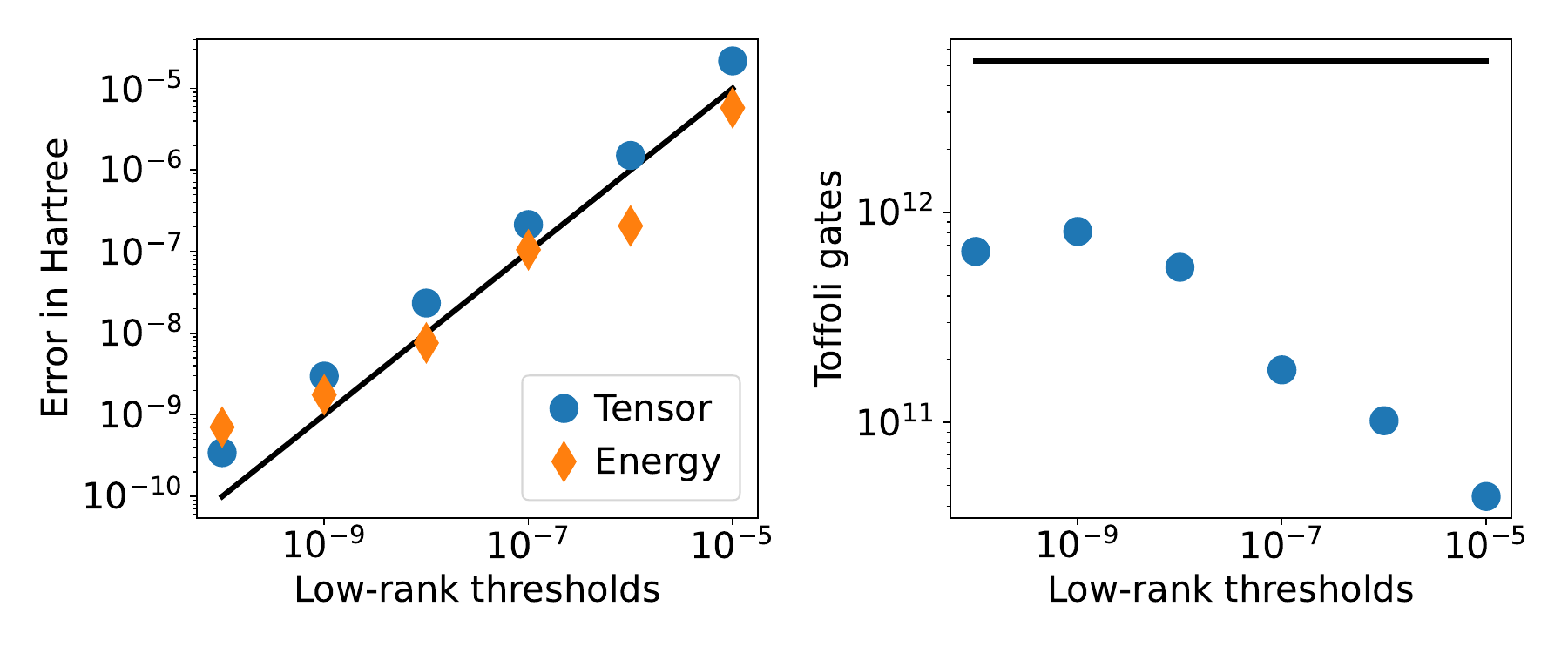}
        \caption{Tensor decomposition errors and gate counts for \ce{H2CO} at the three-mode-coupled level for different low-rank thresholds using five modals per mode. Left figure: tensor and energy errors as a function of low-rank thresholds.
        %for which the horizontal line corresponds to chemical accuracy. 
        Right figure: gate counts for different low-rank thresholds with the cost of the original Hamiltonian indicated by the black line.}
        \label{fig:h2co_low_rank_computation}
    \end{figure}

    For the case of formaldehyde, the tensor and energy error are seen in the left panel of \cref{fig:h2co_low_rank_computation} across a variety of low-rank thresholds.
    A black $y=x$ line makes the errors easily comparable to the low-rank thresholds.
    The right panel shows the Toffoli cost at the same low-rank thresholds along with a black line indicating the Toffoli cost of the original Hamiltonian.
    
    It is observed that the low-rank threshold effectively controls both the tensor and energy error, both of which are comparable in magnitude to the low-rank threshold.
    This indicates that errors introduced by tensor decomposition are mainly stochastic in nature and not systematic.
    %The size of systematic errors scales linearly with the number of terms in the Hamiltonian,
    %which would be detrimental to the scheme.

    The right panel shows a decrease in the Toffoli gate cost by up to two orders of magnitude compared to the exact Hamiltonian.
    Chemical accuracy is obtained at $\epsilon_\textsc{lr} \leq \SI{4.5e-6}{\hartree}$,
    hence the achievable reduction is close to two orders of magnitude.

    \begin{figure}
        \centering
        \includegraphics[width=\columnwidth]{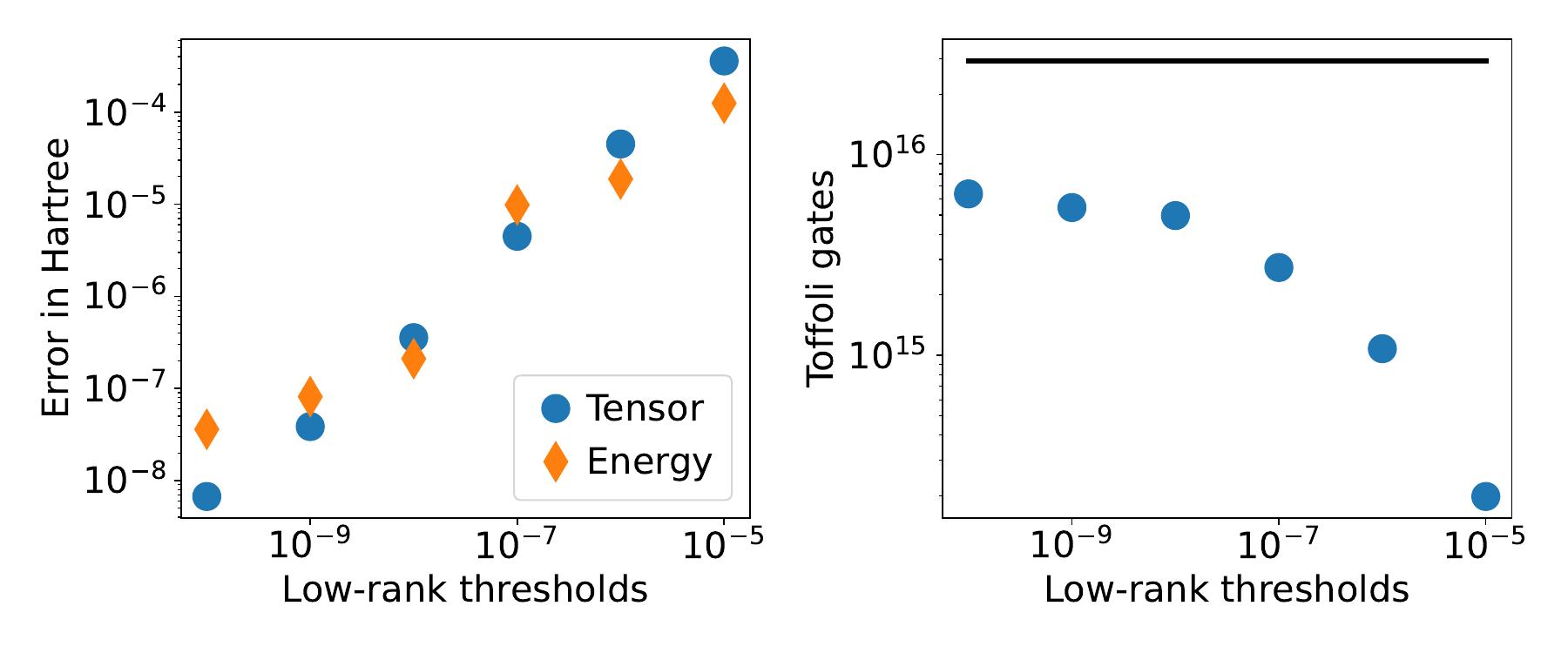}
        \caption{Tensor decomposition errors and gate counts for PAH2 at the three-mode-coupled level for different low rank thresholds using five modals per mode. Left figure: tensor and energy errors as a function of low rank thresholds. 
        Right figure: gate counts for different low-rank thresholds with the cost of the original Hamiltonian indicated by the black line.}
        \label{fig:pah2_3m_low_rank_computation}
    \end{figure}
    
    %The \ac{qpe} cost, $\Delta E(\epsilon_\textsc{t},\epsilon_\textsc{lr})$ and $\epsilon_\textup{c}(\epsilon_\textsc{t},\epsilon_\textsc{lr})$ for PAH2 are presented in \cref{fig:pah2_3m_low_rank_computation} for different values of $\epsilon_\textsc{lr}$.
    Analogous to \cref{fig:h2co_low_rank_computation}, \cref{fig:pah2_3m_low_rank_computation} shows the tensor and energy errors along with the Toffoli cost across a range of low-rank thresholds for PAH2 (naphthalene).
    In terms of gate cost reduction, the tensor decomposition performs comparable to \ce{H2CO}.
    Contrary the formaldehyde, some degree of error compounding is observed and tighter tensor decomposition thresholds are thus required to achieve chemical accuracy.
    One may thus expect a reduction somewhere between one and two orders of magnitude when employing the low-rank factorisation scheme presented here.

    \section{Conclusion}

    In this work, we have considered a variety of algorithms to reduce the circuit depths of quantum computations for vibrational wave functions using qubitization.
    
    For efficient block encoding of vibrational Hamiltonians, we have provided explicit quantum circuits for encoding \ac{sop} operators and methods for optimizing parallel block encodings using different grouping algorithms. These methods proved to reduce circuit depths with up to around two orders of magnitude for large chains of \aclp{pah} between proposed algorithms.
    
    To reduce the number of terms in vibrational Hamiltonians, we presented high order tensor decomposition to obtain low rank approximations using \ac{cp} decompositions. The tensor decomposition methods proved to reduce the number of terms in the vibrational Hamiltonians between one and two orders of magnitude across the benchmark molecules. In addition, we empirically investigated the numerical stability of the tensor decomposition in terms of energy errors.
    
    In order to allow for realistic \ac{pes} with complicated vibrations, we discussed polyspherical coordinate systems and efficient decompositions of the \ac{keo} to reduce circuit depths for block encoding. We investigated the dependency of the maximum mode coupling of \ac{pes}, demonstrating that the number of terms and couplings in the \ac{pes} significantly impacts circuit depths.
    
    \begin{acknowledgments}
    
    O.C. acknowledges support from the Independent Research Fund Denmark through grant number 1026-00122B. I.S. acknowledges support from the Innovation Fund Denmark. The authors acknowledge funding from the Novo Nordisk Foundation through grant NNF20OC0065479. Calculations using MidasCpp (\ac{pes}, tensor decomposition and FVCI/VCC) were performed
    at the Centre for Scientific Computing Aarhus (CSCAA).
    This work was funded by the European Innovation Council through Accelerator grant no. 190124924.
    
    \end{acknowledgments}

    \begin{conflicts}
    The authors have no conflicts to disclose
    \end{conflicts}

    \begin{dataav}
    The code for the new tensor decomposition of the Hamiltonian is available
in the newest MidasCpp release (\href{https://source.coderefinery.org/midascpp/midascpp/-/releases/2025.07.0}{https://source.coderefinery.org/midascpp/midascpp/-/releases/2025.07.0}), which also includes the PES construction methods used. All intermediate results are available from the authors upon reasonable request. 
\end{dataav}

    \appendix
    
    \section{\Acs{qpe} costs} \label{app:sec:qpe_cost}
        
    The cost of \ac{qpe} depends on the cost of the initial state preparation, the encoding of the Hamiltonian, the state preparation of the readout qubits and the inverse \ac{qft}.
    
    The initial state preparation may be carried out using the \ac{vscf} state which may be prepared with negligible cost similarly to the \ac{hf} state in electronic structure. State preparation for the $\ket{0}^{\otimes m}$ readout qubits may similarly be neglected along with the inverse \ac{qft}, which scale as $\mathcal{O}(m)$.
    
    Ignoring the costs of preparing the initial state, the gate cost of \ac{qpe} to reach an overall precision $\epsilon$ is given by~\cite{babbush_encoding_2018}
    \begin{align} \label{eq:qpe_basic_cost}
        \mathcal{C}[\textup{QPE}] = n_\textup{walk} \mathcal{C}[W] = n_\textup{walk} \left( \mathcal{C} \big[ \mathcal{B} [ H ] \big] + \mathcal{C}[R] \right),
    \end{align}
    where $n_\textup{walk}(\epsilon)$ is the number of queries to $W$.
    
    The Toffoli cost of the reflection operator is bounded by the number of qubits the operator acts on, i.e. the number of encoding qubits~\cite{von_burg_quantum_2021}, hence
    \begin{align} \label{eq:reflection_cost}
        \mathcal{C}[R] \leq \mathcal{N}_\textup{enc}.
    \end{align}
    %
    %The number of encoding qubits needed for the block encoding, 
    $\mathcal{N}_\textup{enc}$ is the maximum number of encoding qubits in active use at any one time.
    The number of queries to the walk operator oracle is bounded by $n_\textup{walk} < \sqrt{2} \pi \alpha / \epsilon$~\cite{babbush_encoding_2018}, hence the cost of \ac{qpe}, \cref{eq:qpe_basic_cost}, has the upper bound
    \begin{align} \label{eq:cost_qubitization}
        \mathcal{C}[\textup{QPE}] < \frac{\sqrt{2} \pi \alpha}{\epsilon} \Big( \mathcal{C} \big[ \mathcal{B} [ H ] \big] + \mathcal{N}_\textup{enc} \Big).
    \end{align}

    The number of readout qubits is bounded by~\cite{babbush_encoding_2018}
    \begin{align} \label{eq:readout_qubits}
        \mathcal{N}_\textup{readout} \leq \ceil*{ \log_{2} \left( \frac{\sqrt{2}\pi\alpha}{2\epsilon} \right)}
    \end{align}
    to ensure a precision $\epsilon$,
    which is consistent with \cref{eq:cost_qubitization}.

    \section{One-mode operator costs} \label{app:one-mode_operator}
    This appendix derives the Toffoli and encoding qubit cost function for a one-mode operator, $H_{1\textsc{m}}$ in \cref{eq:factorised_hamiltonian}, with all three representations; quadratic (\cref{eq:one_mode_operator_quad}), triangular (\cref{eq:one_mode_operator_tri}) and diagonal (\cref{eq:one_mode_operator_diag}). \cref{sec:select_implementation,sec:prepare_implementation} describes the circuits used along with their qubit and Toffoli gate costs. The costs are stated in terms of the number of operator coefficients encoded during \textsc{prepare} and the number of operators applied during \textsc{select} respectively, i.e. the number of unitaries in the \ac{lcu} denoted $N$. Additionally, \cref{sec:select_basis_tranf_cost} outlines the cost functions for \textsc{select} when basis-transforming unitaries are employed. In \cref{app:cost_functions} these cost functions are evaluated in order to yield the total one-mode operator cost functions.

    Let $H=\sum_{i}c_{i}U_{i}$ denote a \acl{lcu} (\ac{lcu}) were $U_{i}$ is a unitary operator. The \textsc{select} operator may be constructed such that
    \begin{align}
        S [H] = \sum_{ \mathclap{i=0} }^{ \mathclap{N-1} } \ket{i}_\textup{enc} \prescript{}{\textup{enc}}{\bra{i}} \otimes  U_i
    \end{align}
    where the \textsc{prepare} operator will be defined as
    \begin{align} \label{eq:prepare_operator}
        P[H] = \sum_{ \mathclap{i=0} }^{ \mathclap{N - 1} } \sqrt{ \frac{c_{i}}{\alpha} } \ket{i}_\textup{enc} \prescript{}{\textup{enc}}{\bra{0}}.
    \end{align}
    Here $N$ is the number of unitaries in the \ac{lcu} while the qubits $\ket{0}_\textup{enc}$ and $\ket{i}_\textup{enc}$ are encoding qubits that store a unary encoded integer. The \ac{lcu} norm $\alpha$ is defined in \cref{eq:lcu_norm}.
    In practice, \textsc{prepare} will be implemented with the encoding register entangled to an ancillary register in some unspecified garbage state that will be uncomputed by $\textsc{prepare}\dag$, i.e.
    \begin{align}
        P[H] \ket{0}_\textup{enc} \ket{0}_\textup{anc} = \sum_{ \mathclap{i=0} }^{ \mathclap{N - 1} } \sqrt{ \frac{c_{i}}{\alpha} } \ket{i}_\textup{enc} \ket{\textup{garbage}_i}_\textup{anc}.
    \end{align}

    The Toffoli complexity of the block encoding operator can be broken down to a composition of costs of the \textsc{prepare} and \textsc{select} subroutines. Thus, the cost of the block encoding operator may be written as
    \begin{align} \label{eq:block_encoding_decomposition}
        \mathcal{C} \big[ \mathcal{B} [ H ] \big] = \mathcal{C} [P\dag] + \mathcal{C} [S] + [P]
    \end{align}
    
    \subsection{\textsc{select} implementation} \label{sec:select_implementation}
    Using the unary iteration method presented in Ref.~\cite{babbush_encoding_2018} to perform the multiplex operation, \textsc{select} has an additive cost in the number of terms in the \ac{lcu} and the cost of the unitaries. The cost of the multiplex operation is given by $\mathcal{C}_\textsc{mu}(N) = N - 1$, where $N$ is the number of indices being multiplexed~\cite{babbush_encoding_2018,berry_qubitization_2019}. The diagonal representation requires basis rotating unitaries ($\tilde{U}_i$) during \textsc{select}, whose Toffoli cost adds to that of the multiplex. Thus, the total cost of the \textsc{select} may be written as
    \begin{align} \label{eq:select_cost}
        \mathcal{C} \big[ S [ H_{1\textsc{m}} ] \big] (N) = \mathcal{C}_\textsc{mu}(N) + \sum_{i} \mathcal{C} [ \tilde{U}_{i} ],
    \end{align}
    where $\mathcal{C} [ \tilde{U}_{i} ]$ is the cost of the $i^\textup{th}$ basis-transforming unitary. $i$ is just a dummy index enumerating all basis-transforming unitaries. In case of the quadratic and triangular representations, \cref{eq:select_cost} is still valid with $i \in \emptyset$.

    \subsection{\textsc{prepare} implementation}
    \label{sec:prepare_implementation}
    
    \subsubsection{Coherent alias sampling}
    
    The \textsc{prepare} operator may be implemented using coherent alias sampling as described in Ref.~\cite{babbush_encoding_2018}. The method of coherent alias sampling encodes the coefficients of the \ac{lcu} from a uniform distribution. Thus, the method requires the preparation of a uniform distribution across $N$ states, a data lookup, a comparator and a controlled \textsc{swap} operation.

    Define $\mu$ as the number of qubits used to store one \ac{lcu} coefficient, which may be written as~\cite{babbush_encoding_2018}
    \begin{align} \label{eq:lcu_coef_precision}
        \mu = \ceil*{ \log_{2} \left( \frac{2\sqrt{2}\alpha}{\epsilon} \right) }.
    \end{align}
    $\epsilon$ is defined in \cref{app:sec:qpe_cost}.
    The Toffoli costs of the coherent alias sampling method with $N$ elements are
    \begin{align} \label{eq:data_lookup_babbush}
        \mathcal{C}_\textsc{d}(N) = N - 1
    \end{align}
    for the data lookup,
    \begin{align} \label{eq:comparator_cost}
        \mathcal{C}_\textsc{c}(\mu) = 2\mu - 1
    \end{align}
    for the comparator and
    \begin{align} \label{eq:swap_cost}
        \mathcal{C}_\textsc{s}(N) = \ceil{\log_2(N)}
    \end{align}
    for the controlled \textsc{swap} operation.
    
    The preparation of the uniform superposition requires the preparation of a uniform distribution across $N$ states. A data look-up requires an encoding register of size
    \begin{align} \label{eq:encoding_qubits_prepare}
        \mathcal{N}_\textup{enc}(N) = \ceil{\log_2(N)}
    \end{align}
    and a size-$n_\textup{bits}$ value register to store the values of the look-ups. Additionally, an ancilla qubit is needed for each encoding qubit in a data look-up.
    %In the case of $\mu$-bit approximations to the \ac{lcu} coefficients $n_\textup{bits} = \mu$.
    
    There are two ways to perform the uniform state preparation.
    First, using $\log_{2}(N)$ qubits, the state preparation requires amplitude amplification when $N$ is not a power of two. Using the method in Ref.~\cite{babbush_encoding_2018}, the state preparation may be performed using Hadamard gates followed by amplitude amplification that realizes the desired state. However, the costs of amplitude amplification are not negligible and may result in infeasible overheads.
    
    Second, in Ref.~\cite{von_burg_quantum_2021}, a different method was proposed requiring at most twice as many Toffoli gates for the data lookups while requiring only Hadamard gates for preparing the uniform superposition. Let $\tilde{N}=\ceil*{\log_{2}(N)}$ denote the number of qubits required to encode the set of coefficients $S=\{c_{i}|0\leq i \leq N-1\}$ and $S_{0}=\{0|N\leq i \leq \tilde{N} - 1\}$ such that $|S|+|S_{0}|=2^{\tilde{N}}$. With this method, the number of coefficients for the data lookup are at most twice as large as $N$. To mitigate the overheads for performing the amplitude amplification in the first method, we utilize the second method in this work.

    The total Toffoli cost of the \textsc{prepare} operator using the coherent alias sampling method may then be written as
    \begin{align} \label{eq:general_prepare_toffoli}
        \mathcal{C} [P] = \mathcal{C}_\textsc{d}(N)+ \mathcal{C}_\textsc{c}(\mu) + \mathcal{C}_\textsc{s}(N).
    \end{align}
    
    The \textsc{prepare} subroutine must load two databases of sizes $\ceil{\log_2(N)}$ and $\mu$ respectively during the same multiplex operation, $n_\textup{bits} = \ceil{\log_2(N)} + \mu$. Additionally, a $\mu$-bit register, along with a single comparator result qubit, is needed. With the method described in Ref.~\cite{babbush_encoding_2018} these serve an ancillary function, hence the circuit requires
    \begin{align} \label{eq:general_prepare_qubit}
        \mathcal{N}_\textup{anc}[P] = \ceil{\log_2(N)} + 2 \mu + 1
    \end{align}
    ancillary qubits along with the encoding qubits in \cref{eq:encoding_qubits_prepare}. The ancillary register needed for the data look-up, we choose to count separately as
    \begin{align} \label{eq:babbush_data_lookup_ancillae}
        \mathcal{N}_\textup{clean}[P] = \ceil{\log_2(N)}
    \end{align}
    for a reason that will become apparent in \cref{app:cost_functions}.

    Note that using the standard coherent alias sampling technique the cost functions for computing and uncomputing ($P$ and $P\dag$) are identical.

    \subsubsection{Toffoli and qubit trade-offs} \label{sec:clean-dirty_commpute}
    
    In Refs.~\cite{low_trading_2018,berry_qubitization_2019}, data lookup methods are described where Toffoli gates are traded for additional qubits achieving sublinear state preparation complexity. One may utilize $\lambda_\textsc{c} > 1$ registers, with $\lambda_\textsc{c}$ being a power of $2$, to reduce the Toffoli costs. One of the registers will be used as the output register, with the remaining $\lambda_\textsc{c} - 1$ registers being ancillary. Note that this subsection only reports the extra ancillary qubits required for the data look-up, and not the encoding and data register sizes.

    Recall from the main text, \cref{sec:one-mode_operators}, that we refer to qubits whose initial state is arbitrary as dirty and those that must be initialised to $\ket{0}$, as clean.
    In both cases the qubits are returned to their initial state at the end.
    Using the auxillary variable $a$ to take the value $a = 1$ when strictly using clean qubits and $a = 2$ when dirty qubits are allowed, the cost functions for data look-ups can be stated concisely.
    Using clean qubits only, the Toffoli cost of data look-ups may be reduced from $\mathcal{C}_\textsc{d}(N) = N - 1$ to
    \begin{align} \label{eq:clean_dirty_data_lookup}
        \mathcal{C}_\textsc{d,c} (N, n_\textup{bits}, \lambda_\textsc{c}) &= a\ceil*{\frac{N}{\lambda_\textsc{c}}} + a^2 n_\textup{bits} (\lambda_\textsc{c} - 1).
        \intertext{This requires}
        \mathcal{N}_\textup{clean} (N, n_\textup{bits}, \lambda_\textsc{c}) &= \ceil*{\log_{2} \left( \frac{N}{\lambda_\textsc{c}} \right)} \nonumber \\
        &+ (2 - a) n_\textup{bits}(\lambda_\textsc{c} - 1) \label{eq:clean_qubits}
        \intertext{clean, ancilla qubits and}
        \mathcal{N}_\textup{dirty} (n_\textup{bits}, \lambda_\textsc{c}) &= (a - 1) n_\textup{bits} (\lambda_\textsc{c} - 1) \label{eq:dirty_qubits}
    \end{align}
    dirty, ancillary qubits.
    
    Notice that the number of dirty qubits is independent of the size of the database and the number of clean qubits is independent on the floating point precision.
    The parameter $\lambda_\textsc{c}$ may be optimized to yield the lowest Toffoli cost.
    The optimum occurs at
    \begin{align} \label{eq:optimal_lambda_compute}
        \lambda_\textsc{c} \sim \lambda_\textsc{c}^\textup{opt} = \sqrt{\frac{aN}{n_\textup{bits}}}.
    \end{align}
    %
    %using the auxillary variable $a$ to take the value $a=1$ when strictly using clean qubits and $a=2$ when dirty qubits are allowed.

    \subsubsection{Measurement-based uncomputation} \label{sec:clean-dirty_uncommpute}
    
    In addition to the trade-off between Toffoli gates and qubits, the Toffoli cost of uncomputing the state preparation may be further reduced using measurement-based uncomputation, as described in Ref.~\cite{berry_qubitization_2019}. Importantly, the associated cost is independent of the size of the database and $\lambda_\textsc{u}$ is a number of qubits rather than a number of registers. $\lambda_\textsc{u} > 1$ must also be a power of $2$. Additionally, $\lambda_\textsc{u}$ may also be different from $\lambda_\textsc{c}$, and they may thus be optimised indepently.

    Measurement-based uncomputation requires
    \begin{align}
        \mathcal{C}_\textsc{d,u} (N,\lambda_\textsc{u}) = a \ceil*{ \frac{N}{\lambda_\textsc{u}} } + a^2 \lambda_\textsc{u} \label{eq:clean_dirty_uncompute}
    \end{align}
    Toffoli gates,
    where the optimal value of $\lambda_\textsc{u}$ is
    \begin{align} \label{eq:optimal_lambda_uncompute}
        \lambda_\textsc{u} \sim\lambda_\textsc{u}^\textup{opt} = \sqrt{\frac{N}{a}}.
    \end{align}
    Because $\lambda_\textsc{c}$ is a number of registers and $\lambda_\textsc{u}$ is a number of qubits, computing a data look-up will in practice require more ancilla qubits than the following uncomputation.
    No additional qubits are therefore needed for the measurement-based uncomputation protocol.

    \subsection{\textsc{select} with basis-transforming unitaries} \label{sec:select_basis_tranf_cost}
    Care must be taken when implementing operators containing basis-transforming unitaries, as these turn out to be expensive.
    We choose to implement \textsc{select} of the first term in the diagonal representation of a one-mode Hamiltonian, \cref{eq:one_mode_operator_diag}, as
    \begin{align} \label{eq:select_block_encode_step}
    \begin{aligned}
        S \Big[ \sum_{ \mathclap{j^m\alpha} } \tau_{j^m}^{\alpha} \upsilon_{j^m}^{\alpha} \Big] = &\sum_{j^m} \ket{j^m}\bra{j^m} \\
        &\otimes \tilde{U}_p^{\dagger} \mathcal{B} \Big[ \sum_\alpha \sigma_{0}^{\alpha} \tilde{U}_p \tilde{U}_q\dag \sigma_{0}^{\alpha} \Big] \tilde{U}_q.
    \end{aligned}
    \end{align}
    Here $\tilde{U}$ denotes a basis-transforming unitary.
    This way one needs to implement six unitaries, rather than eight, for each term in the modal sum ($j^m$), for a total of $6N_m$ unitaries.

    Additionally, we choose to create a size-$2N_m$ database with elements of size $N_m$, i.e. a database containing $2N_m$ vectors of $N_m$ $\beta$-bit rotation angles. This requires
    \begin{align} \label{eq:select_database_qubits}
        \mathcal{N}_\textup{anc}^\textsc{d}[S] = \ceil{\log_2(2N_m)} + \beta N_m
    \end{align}
    qubits to store and index the rotation angles during \textsc{select}. These rotation angles are read with data look-ups in order to implement the correct basis rotations before the data look-up is uncomputed. Each basis-transforming unitary therefore requires $\mathcal{C}_{\textsc{d},\textsc{c}}(2 N_m, N_m \beta) + \mathcal{C}_{\textsc{d},\textsc{u}}(2 N_m)$ Toffoli gates for data look-ups.

    With the correct rotation angle loaded, the rotation itself also requires Toffoli gates. This may be done using the method derived in \cref{sec:basis_transformation_bosons} which shows that each unitary transformation can be implemented as two $N_m$-fold products of two-qubit rotations with each angle stored as a $\beta$-bit approximation. Each unitary therefore turn out to require $2\beta N_m$ Toffoli gates.

    Lastly, the $\alpha$ sum requires a two-term multiplex. Combining this multiplex with the cost of the database and the controlled rotations, the total Toffoli cost associated with the unitary basis transformation becomes
    \begin{align} \label{eq:select_unitaries_contribution}
    \begin{aligned}
         \sum_{i} \mathcal{C} [ U_{i} ] &= C_{\textsc{d},\textsc{c}}(2 N_m, N_m \beta)+ \mathcal{C}_{\textsc{d},\textsc{u}}(2 N_m, N_m \beta) \\
         &+ N_m \big[ 12 \beta N_m + C_\textsc{mu}(2) \big].
    \end{aligned}
    \end{align}
    Recall that $i$ is just a dummy index enumerating all basis-transforming unitaries. If ancilla qubits are used to speed up the data look-ups this requires
    \begin{subequations}
    \begin{align}
        \mathcal{N}_\textup{clean} &= \mathcal{N}_\textup{clean} (2N_m, N_m \beta, \lambda_\textsc{c}), \\
        \mathcal{N}_\textup{dirty} &= \mathcal{N}_\textup{dirty} (N_m \beta, \lambda_\textsc{c}), 
    \end{align}
    \end{subequations}
    where the value of $\lambda_\textsc{c}$ will be addressed in \cref{app:sec::lambda_value}.

    \subsection{Cost functions} \label{app:cost_functions}
    The unary itarations performed during \textsc{prepare} and \textsc{select} use the same encoding qubits, the number of which is determined by the size of the \ac{lcu} $N$, hence
    \begin{align} \label{eq:general_encoding_qubits}
        \mathcal{N}_\textup{enc} \big[ \mathcal{B} [ H_{1\textsc{m}} ] \big] = \ceil{\lg{N}}.
    \end{align}

    \textsc{prepare} uses extra ancillary qubits, and with the diagonal representation, so does \textsc{select}. As the diagonal representation reads the rotation angles from a database with data look-ups, ancillae are needed for the rotation angle values and indices, \cref{eq:select_database_qubits}.

    By counting the ancillae used for data look-ups with the symbol $\mathcal{N}_\textup{clean}$, as in \cref{eq:babbush_data_lookup_ancillae}, the variables $\mathcal{N}_\textup{enc}$ and $\mathcal{N}_\textup{anc}$ become independent of the data look-up algorithm. Ancilla requirements for the data look-up oracles are thus handled by $\mathcal{N}_\textup{clean}$ and $\mathcal{N}_\textup{dirty}$ with no extra bookkeeping needed.
    
    \cref{sec:clean-dirty_commpute,sec:clean-dirty_uncommpute} show that uncomputing data look-ups can be done using fewer ancillary qubits than the data loop-up itself, hence $P\dag$ never requires more ancillae that $P$. The number of ancillae needed to block encode a one-mode operator is the largest number required at any point in time,
    \begin{align} \label{eq:general_encoding _qubits}
        \mathcal{N}_\textup{anc} \big[ \mathcal{B} [ H_{1\textsc{m}} ] \big] = \max \left( \mathcal{N}_\textup{anc} [ P ], \mathcal{N}_\textup{anc} [ S ] \right).
    \end{align}

    \subsubsection{Quadratic and triangular representations} \label{sec:quadratic_triangular_cost_app}
    
    Using the results from the previous subsections, \cref{eq:data_lookup_babbush,eq:comparator_cost,eq:swap_cost,eq:general_prepare_toffoli},
    \textsc{prepare} and its inverse can be implemented using
    \begin{align} \label{eq:quadratic_plain_prepare_toffoli}
        \mathcal{C} [ P ] = \mathcal{C} [ P\dag ] &= N + \ceil{\log_2 (N)} + 2\mu - 2
    \end{align}
    Toffoli gates.
    This yields a total of
    \begin{align} \label{eq:toffoli_cost_plain_quadratic}
        \mathcal{C} [ \mathcal{B} ] &= 3N + 2 \ceil{\log_2(N)} + 4\mu - 5.
    \end{align}
    Toffoli gates to block encode a one-mode operator with the quadratic representation.
    If one chooses to use qubits in order to lower the Toffoli gate depth, the cost functions for \textsc{prepare} become
    \begin{subequations}
    \begin{align}
    \begin{aligned}
        \mathcal{C} [ P ] &= a \ceil*{\frac{N}{\lambda_\textsc{c}}} + \ceil{\log_2(N) } + 2 \mu \\
        &+ a^2 \left( \ceil*{\lg (N) } + \mu \right) (\lambda_\textsc{c} - 1) - 1
    \end{aligned}
    \end{align}
    and
    \begin{align}
    \begin{aligned}
        \mathcal{C} [ P\dag ] &= a \ceil*{\frac{N}{\lambda_\textsc{u}}} + a^2 \lambda_\textsc{u} + \ceil{\log_2(N) } + 2 \mu - 1
    \end{aligned}
    \end{align}
    \end{subequations}
    using \cref{eq:clean_dirty_data_lookup,eq:clean_dirty_uncompute}.
    This yields a total Toffoli cost of
    \begin{align} \label{eq:one-mode_toffoli_cost_quad_clean_dirty}
    \begin{aligned}
        \mathcal{C} [ \mathcal{B} ] &= N + a \sum_i \ceil*{\frac{N}{\lambda_i}} + 2 \ceil{\log_2(N) } \\
        &+ a^2 \left( \ceil*{\lg (N) } + \mu \right) (\lambda_\textsc{c} - 1) \\
        &+ a^2 \lambda_\textsc{u} + 4\mu - 3.
    \end{aligned}
    \end{align}
    Here $i \in \{ \textsc{c}, \textsc{u} \}$ is an index that succinctly handles that $\lambda_\textsc{c}$ and $\lambda_\textsc{u}$ may be different. Achieving this number of Toffoli gates requires
    \begin{align} \label{eq:quad_ancilla_cost_app}
    \begin{aligned}
        \mathcal{N}_\textup{clean} [ \mathcal{B} ] &= \ceil*{\log_{2} \left( \frac{N}{\lambda_\textsc{c}} \right) } - 1 \\
        &+ (2 - a) \left( \ceil*{\lg (N) } + \mu \right) (\lambda_\textsc{c} - 1)
    \end{aligned}
    \end{align}
    and
    \begin{align}
        \mathcal{N}_\textup{dirty} [ \mathcal{B} ] &= (a - 1) \left( \ceil*{\lg (N) } + \mu \right) (\lambda_\textsc{c} - 1)
    \end{align}
    ancilla qubits, \cref{eq:clean_qubits,eq:dirty_qubits}, specifically for the data look-up oracles.% in addition to \cref{eq:quadratic_prepare_qubit}.
    As \textsc{select} requires fewer ancillae than \textsc{prepare},
    the ancilla requirement is given by \cref{eq:general_prepare_qubit}.

    Using the Iverson bracket notation
    \begin{align} \label{eq:iverson_bracket}
        [x] = \begin{cases}
            1, \enspace x \textup{ is true} \\
            0, \enspace x \textup{ is false} \\
        \end{cases}
    \end{align}
    \cref{eq:toffoli_cost_plain_quadratic,eq:one-mode_toffoli_cost_quad_clean_dirty} be combined in a single equation as
    \begin{align} \label{eq:one-mode_toffoli_cost_quadratic_app}
    \begin{aligned}
        \mathcal{C} [ \mathcal{B} ] &= N \big(1 + 2[a = 0] \big) + 2 \ceil{\log_2(N) } \\
        &+ a \sum_i \ceil*{\frac{N}{\lambda_i}} + a^2  \lambda_\textsc{u} \\
        &+ a^2 \left( \ceil*{\lg (N) } + \mu \right) (\lambda_\textsc{c} - 1) \\
        &+ 4\mu - 3 - 2[a = 0].
    \end{aligned}
    \end{align}
    Here, $a = 0$ if ancillary qubits are not used.

    A one-mode operator in the quadratic representation, \cref{eq:one_mode_operator_quad}, has
    \begin{align} \label{eq:quadratic_lcu_coef_number_app}
        N = N_m^2
    \end{align}
    coefficients stored as $\mu$ bit approximations, hence
    \begin{align} \label{eq:quadratic_prepare_data_lookup_bits}
        n_\textup{bits} = \ceil*{\lg (N_m^2) } + \mu.
    \end{align}
    The triangular representation, \cref{eq:one_mode_operator_tri}, has
    \begin{align} \label{eq:triangular_lcu_coef_number_app}
        N = N_m (N_m + 1)/2
    \end{align}
    coefficients stored at $\mu$ bit approximations, hence
    \begin{align} \label{eq:triangular_prepare_data_lookup_bits}
        n_\textup{bits} = \ceil*{\lg \left(N_m \frac{N_m + 1}{2} \right) } + \mu.
    \end{align}
    Combining \cref{eq:quadratic_lcu_coef_number_app,eq:quadratic_prepare_data_lookup_bits} and \cref{eq:triangular_lcu_coef_number_app,eq:triangular_prepare_data_lookup_bits} respectively with \cref{eq:one-mode_toffoli_cost_quadratic_app,eq:encoding_qubits_prepare,eq:quad_ancilla_cost_app} the asymptotic scalings in \cref{tab:asymptotic_costs_gates,tab:asymptotic_costs_qubits} in the main text are obtained.

    \subsubsection{Diagonal representation} \label{sec:diagonal_cost_app}

    The Toffoli cost function of the diagonal representation is obtained by adding the contribution from the basis-transforming unitaries, \cref{eq:select_unitaries_contribution}, to \cref{eq:one-mode_toffoli_cost_quadratic_app}.
    A one-mode operator in the diagonal representation, \cref{eq:one_mode_operator_tri}, has
    \begin{align} \label{eq:diagonal_lcu_coef_number}
        N = 2N_m
    \end{align}
    $\mu$-bit \ac{lcu} coefficients, hence
    \begin{align}
        n_\textup{bits} = \ceil*{\lg ( 2 N_m ) } + \mu.
    \end{align}
    The cost function thus becomes

    \begin{align} \label{eq:one-mode_toffoli_cost_diag_app}
    \begin{aligned}
        \mathcal{C} [ \mathcal{B} ] &= N_m \big[ 12 \beta N_m + 3 \big] + 2 \ceil{\log_2(N_m)} \\
        &+a \sum_{i,j} \ceil*{\frac{2N_m}{\lambda_i^j}} \\
        &+ a^2 N_m \beta (\lambda_\textsc{c}^\textsc{s} - 1) \\
        &+ a^2 \left( \ceil*{\lg ( 2 N_m ) } + \mu \right) (\lambda_\textsc{c}^\textsc{p} - 1) \\
        &+ a^2 \sum_j \lambda_\textsc{u}^j \\
        &+ 4\mu - 1.
    \end{aligned}
    \end{align}

    Adding the required number of ancillae, \cref{eq:select_database_qubits}, to \cref{eq:general_prepare_qubit} the ancilla count becomes
    \begin{align} \label{eq:plain_prepare_ancilla_qubit_diagonal}
        \mathcal{N}_\textup{anc} [ \mathcal{B} ] &= 2 \ceil{\log_2 ( N_m )} + \beta N_m + 2 \mu + 3.
    \end{align}
    The qubit requirements of the data look-up circuit becomes
    \begin{subequations} \label{eq:diagonal_data_lookup_ancilla}
    \begin{align} \label{eq:diagonal_prepare_clean}
    \begin{aligned}
        \mathcal{N}_\textup{clean} [ P ] &= \ceil*{ \log_2 \left( \frac{N_m}{\lambda_\textsc{c}^\textsc{p}} \right) } + 1 \\
        &+ (2 - a) \left( \ceil*{\lg ( 2 N_m ) } + \mu \right) (\lambda_\textsc{c}^\textsc{p} - 1),
    \end{aligned}
    \end{align}
    \begin{align} \label{eq:diagonal_prepare_dirty}
    \begin{aligned}
        \mathcal{N}_\textup{dirty} [ P ] &= (a - 1) \left( \ceil*{\lg ( 2 N_m ) } + \mu \right) \\
        &\times (\lambda_\textsc{c}^\textsc{p} - 1) + 1,
    \end{aligned}
    \end{align}
    \begin{align} \label{eq:diagonal_select_clean}
    \begin{aligned}
        \mathcal{N}_\textup{clean} [ S ] &= \ceil*{ \log_2 \left( \frac{N_m}{\lambda_\textsc{c}^\textsc{s}} \right) } \\
        &+ (2 - a) \beta N_m (\lambda_\textsc{c}^\textsc{s} - 1) + 1,
    \end{aligned}
    \end{align}
    and
    \begin{align} \label{eq:diagonal_select_dirty}
        \mathcal{N}_\textup{dirty} [ S ] = (a - 1) \beta N_m (\lambda_\textsc{c}^\textsc{s} - 1) + 1,
    \end{align}
    \end{subequations}
    respectively. \cref{eq:one-mode_toffoli_cost_diag_app,eq:diagonal_data_lookup_ancilla,eq:plain_prepare_ancilla_qubit_diagonal,eq:encoding_qubits_prepare} result in the asymptotic scalings in \cref{tab:asymptotic_costs_gates,tab:asymptotic_costs_qubits} in the main text.

    \subsection{Tradeoff parameter selection} \label{app:sec::lambda_value}
    The value of $\lambda_i$, $i \in \{\textsc{c}, \textsc{u} \}$, used in an actual \ac{qpe} calculation will inevitably depend on the number of qubits available on the physical hardware. As the use of dirty qubits will always be more expensive in terms of Toffoli gates than using the same number of clean qubits, the use of dirty qubits is only meaningful in the case where the number of qubits is limited.
    %We choose to study two cases.
    %First, we consider the case where no extra qubits are used and the cost of data look-ups is given by \cref{eq:data_lookup_babbush}.
    
    As a best-case scenario, where an unlimited number of available qubits is assumed, we consider the case where $\lambda_i$ assumes its optimal value strictly using clean qubits. That $\lambda_i$ is indeed a power of 2 is ensured by selecting the power of 2 nearest to the optimum, \cref{eq:optimal_lambda_compute,eq:optimal_lambda_uncompute}. In case the optimum is equidistant from two powers of two, the smaller is chosen, because if the two choices are equal in Toffoli complexity, we want the choice that uses the smallest number of qubits.

    In this best-case scenario an increase in Toffoli cost of more than \SI{10}{\percent} is observed. In particular, the cost of \ce{H2CO} and PAH1 increases by \SI{16}{\percent} and \SI{12}{\percent}, respectively, compared to the results reported in \cref{tab:results}.
    Therefore, the \textsc{selectswap} circuit was deemed inferior in this study.
    
    \section{Basis transformation of bosonic operators} \label{sec:basis_transformation_bosons}
    Consider the basis transformation of a bosonic operator
    \begin{align}
        b_{q} \defeq \left(U^{(rs)}\right)^{\dagger} \anni{}{q} U^{(rs)},
    \end{align}
    where $q, r$ and $s$ are restricted to the same mode.
    The mode indices are suppressed in this subsection for simplicity.
    
    Let
    \begin{align} \label{eq:transformation_generator}
        \kappa_{rs}=\crea{}{r}\anni{}{s}-\crea{}{s}\anni{}{r}
    \end{align}
    denote an anti-Hermitian excitation operator excitation operator between modals $r$ and $s$.
    Note that the excitation operators are restricted to the physical space where only one modal is occupied per mode, meaning the wave function and the result of the Hamiltonian acting on the wavefunction are linear combinations of kets $\ket{\mathbf{k}}=\ket{k_{1}^1, \cdots, k_{r^m}^m, \cdots, k_{N^M}^M}$ where $k_{r^m}^m$ satisfying $\forall m: \sum_{p^m} k_{r^m}^m= 1\}$ where $k_{r^m}^m$. In the physical space, within one mode, $\anni{}{r}\anni{}{s}\ket{\mathbf{k}} = 0$ and $\crea{}{r}\crea{}{s}\ket{\mathbf{k}} = 0$. This has a range of consequences which can be expressed leaving out the $\ket{\mathbf{k}}$ for brevity. First, $\tau_{\mu}^2 = 0$. Second, $\kappa_{rs}\anni{}{q}\kappa_{rs}=0$. Third, $[a_q,\crea{}{r}]=[a_q,\anni{}{r}]=0$. These relations allow an expansion of the unitary exponential of excitation operators such that
    \begin{align}
        U^{(rs)}&=e^{\theta_{rs}\kappa_{rs}}
        \\
        &=1+\sin(\theta_{rs})\kappa_{rs}-(\cos(\theta_{rs})-1)\kappa_{rs}^{2}.
    \end{align}
    which rotates between modals $r$ and $s$.

    Using the identities from above we arrive at
    \begin{align} \label{eq:annex_bosonic_op_exponential}
        b_q &= \left(U^{(rs)}\right)^{\dagger} \anni{}{q} U^{(rs)} \\
            &= \anni{}{q} + \sin(\theta_{rs})[\anni{}{q},\kappa_{rs}] \\
            &- (\cos(\theta_{rs})-1)\{\anni{}{q},\kappa_{rs}^{2}\},
    \end{align}
    and, specifically, when $q \notin \{r,s\}$
    \begin{align}
        b_{q} = \anni{}{q},\quad q \notin \{r,s\}.
    \end{align} 
    On the other hand, when $q=r$ we get
    \begin{subequations}
    \begin{align}
        [\anni{}{r},\kappa_{rs}]= &(1-\delta_{rs})\anni{}{s}, \\
        \{\anni{}{r},\kappa_{rs}^{2}\} = & (\delta_{rs} - 1)\anni{}{r},
    \end{align}
    \end{subequations}
    such that for $r \neq s $
    \begin{align}
        b_q=\cos(\theta_{rs})\anni{}{r} + \sin(\theta_{rs})\anni{}{s}.
    \end{align}
    For the case when $r=s$ it follows from \cref{eq:annex_bosonic_op_exponential} that $b_q= \anni{}{r}$.
    Similarly, for $q=s$ we get
    \begin{subequations}
    \begin{align}
        [\anni{}{s},\kappa_{rs}]= &(\delta_{rs} - 1)\anni{}{r}, \\
        \{\anni{}{s},\kappa_{rs}^{2}\} = &(\delta_{rs} - 1)\anni{}{s},
    \end{align}
    \end{subequations}
    such that for $r\neq s$
    \begin{align}
        b_q =\cos(\theta_{rs})\anni{}{s} - \sin(\theta_{rs})\anni{}{r}.
    \end{align}
    Thus, for $r\neq s$ we have established
    \begin{align}\label{eq:cases}
        b_q =
        \begin{cases}
        \anni{}{q}, & q \notin \{r,s\}, \\
        \cos(\theta_{rs})\anni{}{r} + \sin(\theta_{rs})\anni{}{s} & q = r,\\
        \cos(\theta_{rs})\anni{}{s} - \sin(\theta_{rs})\anni{}{r} & q = s.\\
        \end{cases}
    \end{align}
    
    We can then choose to sort our basis and express the subsequent operators in terms of the previous one with $U^{(r, r+1)} = e^{\theta_r \kappa_{r, r+1}}$ and choosing $\theta_r = \pi/2$. Furthermore, the same strategy can be used to implement arbitrary basis transformations starting from the first operator $a_0$.
    
    Through a series of such transformations one may obtain any basis transformation for each mode such that
    \begin{align}
    \begin{aligned}\label{eq:trafo_def}
        b_q   =~&\tilde{U}_q\dag a_0 \tilde{U}_q\\
        =~&\left(U^{(N_{m}-2,N_{m}-1)}_q \right)^{\dagger} \dots{} \left(U^{(01)}_q \right)^{\dagger} \\ &a_{0}U^{(01)}_q \dots{} U^{(N_{m}-2,N_{m}-1)}_q.
    \end{aligned}
    \end{align}
    Using the trigonometric forms of \cref{eq:cases} the recursive transformation may be written explicitly, starting with the transformation of $\anni{}{0}$ and $\anni{}{1}$ to $\bar{a}_0$ and $\bar{a}_1$,
    \begin{align}
        \bar{a}_1 &= \left(U^{(01)}_q \right)^{\dagger}\anni{}{0}U^{(01)}_q = \cos(\theta_0) \anni{}{0} + \sin(\theta_0) \anni{}{1} \\
        \bar{a}_2 &= \left(U^{(12)}_q \right)^{\dagger}\bar{a}_{1}U^{(12)}_q \nonumber \\ &= \sin(\theta_0) \left(U^{(12)}_q \right)^{\dagger}a_{1}U^{(12)}_q + \cos(\theta_0) a_0 \nonumber \\
        &= \sin(\theta_0) \big[ \cos(\theta_1) a_1 + \sin(\theta_1) a_2 \big] + \cos(\theta_0) a_0, \\
        \dots
    \end{align}
    where $\bar{a}_i$ are partially transformed intermediate components in the construction of $\tilde{a}_q$. In fact, both $\theta_i$ and $\bar{a}_i$ depend on the target $b_q$, but the annotation has been omitted for simplicity.\\

    Generally, we can write
    \begin{align}
    \begin{aligned}
        \bar{a}_p = &  \left(U^{(p-1,p)}_q \right)^{\dagger}\bar{a}_{p-1}U^{(p-1,p)}_q \\
                =& \prod_{ \mathclap{i=0} }^{ \mathclap{p-2}} \sin(\theta_i) \left(U^{(p-1,p)}_q \right)^{\dagger}a_{p-1}U^{(p-1,p)}_q \\
                &+ \sum_{i=0}^{p-2}\prod_{j=0}^{i-1} \sin(\theta_j) \cos(\theta_i) a_i \\
                =& \prod_{ \mathclap{i=0} }^{ \mathclap{p-2}} \sin(\theta_i) [\cos(\theta_{p-1}) a_{p-1}+ \sin(\theta_{p-1}) a_p] \\
                &+ \sum_{i=0}^{p-2}\prod_{j=0}^{i-1} \sin(\theta_j) \cos(\theta_i) a_i,
    \end{aligned}
    \end{align}
    which fixes the parameters for each annihilation operator. We can then write the basis transformation
    \begin{align}
        b_q = \sum_i u_i a_i,
    \end{align}
    using the parameters
    \begin{align} \label{eq:trafo_expansion_coeffs}
        u_i = \prod_j^{i-1} \sin(\theta_j) \cos(\theta_i).
    \end{align}
    Specifically, in the qubit representation, the transformation becomes
    \begin{align}\label{eq:successive_unitary_transformation_matrix}
    \begin{aligned}
        U^{(r, r+1)}_q &= e^{2i\theta_r(\sigma_r^x\sigma_{r+1}^y - \sigma_r^y \sigma_{r+1}^x) } \\
        &= e^{2i\theta_r\sigma_r^x\sigma_{r+1}^y } e^{- 2i\theta_r\sigma_r^y \sigma_{r+1}^x}
    \end{aligned}
    \end{align}
    where each of the exponentials may be implemented separately in the same way as in the supplementary material of Ref.~\cite{von_burg_quantum_2021}, Eqns.~(60-63). In contrast to Ref.~\cite{von_burg_quantum_2021}, the starting point of the transformation is $\sigma^\alpha_0$ for both similarity transformations, which we can express as
    \begin{align} \label{eq:off_diagonal_operator}
     \tau_{p}^{\alpha} \upsilon_{q}^{\alpha} = \tilde{U}_p^{\dagger} \sigma_{0}^{\alpha} \tilde{U}_p \tilde{U}^{\dagger}_q \sigma_{0}^{\alpha} \tilde{U}_q,
    \end{align}
    where we have used the definion of $\tilde{U}$ from Eq.~\eqref{eq:trafo_def}. The required rotation angles for these successive transformations can be determined from ~\cref{eq:pauli_transformed,eq:trafo_expansion_coeffs}.

    \subsection{Toffoli cost of basis transformations}
    Suppose the basis transformation operator is dependent on an external index, e.g. as in Eq.~\eqref{eq:one_mode_operator_diag}. Thus, the implementation of a multiplexed transformation over this external index $k = 0, \dots{} K - 1$ for a given step $r, r+1$ in the transformation may be written as
    \begin{align} \label{eq:arbitrary_unitary_rotation}
    \begin{aligned}
        U =&  \sum_{ \mathclap{k=0} }^{ \mathclap{K-1} }  \ket{k} \bra{k} \otimes  U_{q(k)}^{(r, r+1)} \\
        = & \sum_{ \mathclap{k=0} }^{ \mathclap{K-1} } \ket{k} \bra{k} \otimes  e^{2i\theta_r^{(k)} \sigma_r^x\sigma_{r+1}^y } e^{- 2i\theta_r^{(k)}\sigma_r^y \sigma_{r+1}^x}\\
        = & \sum_{ \mathclap{k=0} }^{ \mathclap{K-1} }  \ket{k} \bra{k} \otimes  V_r(\theta_r^{(k)}) W_r(\theta_r^{(j)}),
        \end{aligned}
    \end{align}
    where the notation of Eq.~\eqref{eq:successive_unitary_transformation_matrix} has been extended to take that external index explicitly into account. In addition, $V_r$ and $W_r$ have been implicitly defined. The implementation of this unitary requires a diagonalization of each $V_r$ and $W_r$ followed by controlled rotations
    \begin{align} \label{eq:arbitrary_rotation}
        e^{\pm2i\theta^{(k)}_r \sigma^z_r\sigma_{r+1}^z},
    \end{align}
    where we perform this transformation with $+$ for $V_r$ and $-$ for $W_r$.
    Each term in the sum thus requires two rotations, $V_r$ and $W_r$,
    hence the total number of rotations is
    \begin{align} \label{app:eq:number_of_rotations}
       n_\textup{rot} = 2K.
    \end{align}
    Note that the diagonalization of $V_r$ and $W_r$ may be performed using only Clifford gates and thus their costs are ignored.
    
    The controlled rotations may be implemented using the method presented in Ref.~\cite{von_burg_quantum_2021}. First, the rotation angles $\theta_r^{(k)}$ are loaded as $\beta$-bit approximations from a data lookup oracle. Second, the controlled rotations may be performed using the using the phase gradient technique~\cite{gidney_halving_2018} each with a cost of 1 Toffoli gate. Let $N_m$ denote the number of modals in the current mode. Thus, the total cost of a basis transformation may be written as
    \begin{align}
        \mathcal{C}[U] = \mathcal{C}_\textsc{d}[\theta^{(k)}] + 2\beta N_{m}
        \label{eq:basis_transformation_cost}
    \end{align}
    as there are in total $N_m$ such successive transformation steps and since both $V_r$ and $W_r$ require the application of $\beta$ $k$-multiplexed rotations on qubits $r$ and $r+1$.  The data lookup methods in \cref{sec:prepare_implementation} may be utilized to perform the data lookups for the rotation angles.

    \subsection{Rotation angle precision}
    The rotation angles are loaded as $\beta$-bit approximations, hence an appropriate value of $\beta$ is needed.
    The supplementary material of Ref.~\cite{von_burg_quantum_2021} (Sec. VII 3) shows
    \begin{align} \label{eq:beta_bound}
        \beta = \ceil*{\frac{1}{2} + \log_2 \left( \frac{N\pi}{\epsilon_\textup{rot}} \right) }
    \end{align}
    to be an upper bound on the number of bits needed to ensure that $N$ rotations are implemented to an accuracy $\epsilon_\textup{rot}$.
    \cref{eq:beta_bound} is the result of bounding the error of any single rotation to $\epsilon' = \epsilon_\textup{rot}/N$ using the triangle inequality such that the worst case error is $\epsilon_\textup{rot}$.

    For the current purpose, $n_{\textup{rot}} = 2N_m$ basis transformations per one-mode operator is needed.
    Additionally, the number of modals in mode $m$, $N_m$, may not be the same for all modes. The total number of basis transformations is thus
    \begin{align}
        N_\textup{rot}  &= \sum_{\mathbf{m}} \sum_{t^{\mathbf{m}}} \sum_{ \mathclap{ m \in \mathbf{m} }} n_{\textup{rot}} \\
                        &= 2 \sum_{\mathbf{m}} \sum_{t^{\mathbf{m}}} \sum_{ \mathclap{ m \in \mathbf{m} }} N_m
    \end{align}
    Recall that the purpose of tensor decomposition is to shrink the size of the $t^{\mathbf{m}}$-sum.
    If sufficiently effective, tensor decomposition may thus significantly decrease the number of basis transformations and thereby the required precision of each rotation.
    This in turn decreases the required number of qubits and reduces the Toffoli-depth of the algorithm.

    \subsection{Error budgeting} \label{app:sec:error_budget}
    To achieve an accuracy $\epsilon$ on the final energy estimate, one must ensure that finite precision errors are bounded by $\epsilon$. In addition to the rotation angles, the \ac{lcu} coefficients also contribute finite precision errors. Recall from \cref{eq:lcu_coef_precision} that to ensure finite precision errors from the \ac{lcu} coefficients are kept below $\epsilon_\textsc{lcu}$ the number of qubits required for each coefficient is
    \begin{align}
        \mu = \ceil*{ \log_{2} \left( \frac{2\sqrt{2}\lambda}{\epsilon_\textsc{lcu}} \right) }.
    \end{align}
    An overall accuracy $\epsilon$ is achieved if
    \begin{align} \label{eq:finite_precision_errors}
        \epsilon_\textsc{lcu} + \epsilon_\textup{rot} \leq \epsilon.
    \end{align}
    This is ensured if one chooses
    \begin{subequations}
    \begin{align}
        \epsilon_\textsc{lcu} &= c_\textsc{lcu} \epsilon, \\
        \epsilon_\textup{rot} &= (1 - c_\textsc{lcu}) \epsilon,
    \end{align}
    \end{subequations}
    where $0 < c_\textsc{lcu} < 1$ is a tuneable hyperparameter.

    One may question if \cref{eq:beta_bound} is a tight bound or if a lower number of qubits is sufficient. Additionally, as both $\mu$ and $\beta$ impact the Toffoli cost in different ways, a benchmark study is likely needed to determine the optimal values of $\mu$ and $\beta$. This is beyond the scope of this study, hence we, like Ref.~\cite{von_burg_quantum_2021}, choose $c_\textsc{lcu} = 1/2$, but acknowledge that further optimisation is likely possible.

    \section{Parallelization of block encodings}

    \label{sec:parallelization_block_encodings}
    
    Let $H=\sum_{i}H_{i}$ denote a sum of $N$ Hermitian operators where $\alpha=||H||_{1}$ and $\alpha_{i}=||H_{i}||_{1}$ such that $\alpha=\sum_{i}\alpha_{i}$. Each operator may by assumption be written as an \ac{lcu}, $H_i = \sum_j c_j U_j$. The block encoding of $H$ may be written as a sum of block encodings such that $\mathcal{B}[\frac{H}{\alpha}]=\sum_{i}\mathcal{B}[\frac{H_{i}}{\alpha_{i}}]$. Assume that $H_i$ and $H_j$ act of different subspaces for all pairs $i,j$, such that $[H_{i},H_{j}] = 0$, hence that all operators can be applied simultaneously.

   Our parallelisation strategy relies on the \textsc{fanout} operation $F$ described in~\cite{boyd_low-overhead_2023} that copies data from one register to $N - 1$ other registers,
    \begin{align}
        F \ket{i} \ket{0}^{N - 1} = \ket{i}^{N}.
    \end{align}
    The \textsc{fanout} operation is implemented with a \textsc{cnot} tree, and therefore it has no Toffoli cost. Specifically, let $\ket{i}^{\otimes \lg N}$ be the encoding register of the sum, hence \textsc{fanout} requires $\lg N$ encoding qubits and $(N - 1) \lg N$ ancillary qubits.

    For each Hermitian $H_i$ allocate some size-$M$ encoding register and let
    \begin{align}
        P_i = \sum_j^M \ket{j}_i \prescript{}{i}{\bra{0}}
    \end{align}
    be the \textsc{prepare} operator for the \ac{lcu} decomposition of $H_i$. As all $P_i$ act on separate registers, they may be applied simultaneously.
    \begin{align}
        \tilde{P} = \left( \sum_{i}\ket{i}_{i} \prescript{}{i}{\bra{0}} \right) F.
    \end{align}
    Here $\ket{i}_{i}$ denotes the $i$'th register in the \textsc{fanout} ancillary register in the binary state $\ket{i}$. The \textsc{select} operator for the parallelized block encoding operator may be written as
    \begin{align}
        S = \sum_{i} \ket{i}_{i} \prescript{}{i}{\bra{i}} \otimes \mathcal{B}[H_{i}]
    \end{align}
    such that the block encoding operator may be written as
    \begin{align}
        \mathcal{B}[H]=\tilde{P}^{\dagger}S\tilde{P}
    \end{align}
    The advantage of the \textsc{select} operator is its parallelization since each term may be applied in parallel due to the commutativity of the terms in $H$.
    
    The transformed state using the \textsc{prepare} operator may be written as
    \begin{align}
        \begin{aligned}
            \tilde{P}\ket{\psi}\ket{0}_{i}\ket{\tilde{0}} &= \sum_{i}\ket{\psi}\ket{i}_{i}^{\otimes N}\ket{\tilde{0}}.
        \end{aligned}
    \end{align}
    The transformed state using the \textsc{select} operator may be written as
    \begin{align}
        \begin{aligned}
        S\sum_{i}\ket{\psi}\ket{i}_{i}^{\otimes N}\ket{\tilde{0}} &= \sum_{i}\mathcal{B}[H_{i}]\ket{\psi}\ket{i}_{i}^{\otimes N}\ket{\tilde{0}}.
        \end{aligned}
    \end{align}
    The inverse \textsc{prepare} transforms the above state to
    \begin{align}
        \begin{aligned}
        \tilde{P}^{\dagger}\sum_{i}\mathcal{B}[H_{i}]\ket{\psi}\ket{i}_{i}^{\otimes N}\ket{\tilde{0}}&=
        \sum_{i}\mathcal{B}[H_{i}]\ket{\psi}\ket{0}_{i}\ket{\tilde{0}}.
        \end{aligned}
    \end{align}
    Through projection of the state $\ket{0}_{i}\ket{\tilde{0}}$, the block encoding may be written as
    \begin{align}
        \begin{aligned}
            \bra{\tilde{0}} \prescript{}{i}{\bra{0}}\Big(\sum_{i}\mathcal{B}[H_{i}]\Big)\ket{0}_{i}\ket{\tilde{0}}
            &= \sum_{i}\frac{H_{i}}{\alpha_{i}}
        \end{aligned}
    \end{align}
    since
    \begin{align}
        \begin{aligned}
            \bra{0}_{i}\mathcal{B}[H_{i}]\ket{0}_{i}
            &= \frac{H_{i}}{\alpha_{i}}.
        \end{aligned}
    \end{align}
    
    \section{Tensor decomposition of the vibrational Hamiltonian} \label{sec:tensor_decomposition_appendix}
    
    \subsection{One-mode operators} \label{sec:one_mode_decomposition}
    The one-mode terms may be contracted to one-mode operators for each mode such that
    \begin{align} \label{eq:one-mode_partial_decomposed_operator}
    \begin{aligned}
        H_1 &= \sum_{m}\sum_{o^{m}} c_{o^{m}}^{m} h_{o^{m}}^{m} = \sum_{m} \tilde{h}^{m} \\
    \end{aligned}
    \end{align}
    where 
     \begin{align} \label{eq:one-mode_matrix_efffective}
     \tilde{h}^{m}=\sum_{o^{m}} c_{o^{m}}^{m} h^{m,o^{m}}.
    \end{align}
    We can thus construct a single effective one-mode operator per mode, 
    defined in terms of a similar sum of integrals
     \begin{align} \label{eq:one-mode_matrix_integrals}
     \tilde{h}^{m}_{r^ms^m}=\sum_{o^{m}} c_{o^{m}}^{m} h^{m,o^m}_{r^ms^m}.
    \end{align}
    The operator above may be then be utilized in combination with any representation; quadratic, triangular or diagonal.

    \subsection{Two-mode operators}
    The real two-mode coupling matrices are factorized using an \ac{svd},
    \begin{align}
        c_{o^{m_1}o^{m_2}}^{m_1m_2} =\sum_{k} \lambda_{k}^{m_1m_2} V_{o^{m_1} k}^{m_1, \{m_1,m_2\}} V_{o^{m_2} k}^{m_2, \{m_1,m_2\}},
    \end{align}
    allowing the introduction of transformed one-mode operators specific for the two modes and pair $\{{m_1,m_2}\}$;
    \begin{align} \label{eq:two_mode_zeta}
    \zeta_{k}^{m_i, \{m_1,m_2\}} &= \sum_{o^{m_i}} V_{o^{m_i} k}^{m_i, \{m_1,m_2\}} h^{m_i,o^{m_i}}.
    \end{align}
    The two-mode terms of the Hamiltonian may be thus written as
    \begin{align}
    \begin{aligned}
        H_{2} &= \sum_{m_1 < m_2} \sum_{o^{m_1}o^{m_2}} c_{o^{m_1}o^{m_2}}^{m_1m_2} \bigotimes_{ \mathclap{ i=1,2 } } h_{o^{m_i}}^{m_i} \\
        &= \sum_{m_1 < m_2} \sum_{k}  \lambda_{k}^{m_1m_2} \bigotimes_{ \mathclap{ i=1,2 } } \zeta_{k}^{m_i}.
    \end{aligned}
    \end{align}
    Approximations to $H_2$ may be introduced by truncating the \ac{svd} to make controlled error-bounded approximations to each two-mode coefficient matrix, i.e. controlling the approximation by a single threshold applied to all mode combinations.

    \subsection{Three- and Higher-Mode Operators}
    
    The three- and higher-mode coupling tensors may be factorized in a similar way. 
    For simplicity, this subsection shows equations for the three-mode case, which trivially generalizes to the many-mode case by the addition of an appropriate number of indices.
    Often, it is attractive to use a two-step process, though the first is optional and not decisive for the final format. The optional step is a Tucker decomposition that reduces each three-mode coefficient tensor $c_{o^{m_1}o^{m_3} o^{m_3}}^{m_1m_2m_3}$ to a smaller core tensor ${\tilde c}_{\tilde{o}^{m_1}\tilde{o}^{m_2} \tilde{o}^{m_3}}^{m_1m_2m_3}$ and a set of rectangular matrices $A_{\tilde{o}^{m_i} o^{m_i}}^{m_i}$;
    \begin{align}
        c_{o^{m_1}o^{m_2}o^{m_3}}^{m_1m_2m_3} = 
        \sum_{ \mathclap{\tilde{o}^{m_1}\tilde{o}^{m_2} \tilde{o }^{m_3}} }
        {\tilde c}_{\tilde{o}^{m_1}\tilde{o}^{m_2} \tilde{o }^{m_3}} ^{m_1m_2m_3}  \bigotimes_{ \mathclap{i=1} }^{3}
        A_{\tilde{o}^{m_i}o^{m_i}}^{m_i}.
    \end{align}        
    Here the indices $\tilde{o}^a$ belong to the reduced space.
    The factorisation can then be completed using a \ac{cp} decomposition of the core tensor which, in analogy to the SVD in the two-mode case, can be written as
    \begin{align}
         {\tilde c}_{\tilde{o}^{m_1}\tilde{o}^{m_2} \tilde{o }^{o}}^{m_1m_2m_3} &= \sum_{k} \lambda_{k}^{m_1m_2m_3} \bigotimes_{i=1}^{3} 
         V_{\tilde{o}^{m_i} k}^{m_i, \{m_1,m_2,m_3\}}.
    \end{align}        
    Here, $\lambda_{k}^{m_1m_2m_3} \in \R$ corresponds to the singular values in the two-mode case and $V_{\tilde{o}^{m_i} k}^{m_i, \{m_1,m_2,m_3\}}$ are the factor matrices --- one for each mode of the particular mode coupling. 
    %$b_{ai}$ is a core-space rank-$1$ tensor. 
    As the mode spaces are distinct, one can define
    \begin{align} \label{eq:tucker_core_factorisation}
        U_{o^{m_i} k}^{m_i, \{m_1,m_2,m_3\}} = 
        \sum_{\tilde{o}^{m_i}}
         V_{\tilde{o}^{m_i} k}^{m_i, \{m_1,m_2,m_3\}}
        %b_{ak} 
        A_{\tilde{o}^{m_i}o^ {m_i}}^{m_i}
    \end{align}
    such that
    \begin{align} \label{eq:three_mode_cp}
        c_{o^{m_1}o^{m_2}o^{m_3}}^{m_1m_2m_3} = \sum_{k} \lambda_{k}^{m_1m_2m_3} \bigotimes_{i=1}^3 
       U_{o^{m_i} k}^{m_i, \{m_1,m_2,m_3\}}.
    \end{align}
    Analogous to \cref{eq:two_mode_zeta}, one may define the transformed one mode operators
    \begin{align} \label{eq:three_mode_factorised_operators}
        \zeta_k^{m_i, {m_1m_2m_3}} = \sum_{o^{m_i}}
        U_{o^{m_i} k}^{m_i, \{m_1,m_2,m_3\}} 
        h^{m_i o^{m_i}}
    \end{align}
    such that three-mode terms of the Hamiltonian factorise to the form
    \begin{align} \label{eq:three_body_factorised}
        H_{3} = \sum_{m_1 < m_2 < m_3} \sum_{k} \lambda_{k}^{m_1m_2m_3} 
        \bigotimes_{k=1}^{3} \zeta_k^{m_i, \{m_1m_2m_3\}},
    \end{align}
    when \cref{eq:tucker_core_factorisation,eq:three_mode_factorised_operators} are combined. Notice that $\zeta_k^{m_i, \{m_1m_2m_3\}}$ is a one-mode operator for mode $m_i$ of the triple ${m_1m_2m_3}$ in analogy to the similar two mode case. 

    The decomposition method for three-mode operators generalizes naturally to $n$-mode operators, allowing for a low-rank approximation of the Hamiltonian across all levels of mode coupling. In this general form, we express the full Hamiltonian as
    \begin{align} \label{eq:tensor_decomposed_hamiltonian}
        H = \sum_{\mathbf{m}} \sum_{k} \bigotimes_{ \mathclap{ m \in \mathbf{m} } } \zeta_k^{m, \mathbf{m}},
    \end{align}
    where the singular values has been absorbed into the one-mode operators.
    \cref{eq:tensor_decomposed_hamiltonian} thus takes the form of \cref{eq:factorised_hamiltonian} with the $k$ sum (unique for each mode-combination) taking the place of the sum of products. 

    The Tucker decompositions are performed using \ac{hooi} where each \ac{svd} is truncated to an error $\epsilon_\textsc{t}$. The \ac{cp} decomposition is performed on the Tucker tensor where the \ac{cp} ranks are truncated to an error $\epsilon_\textsc{lr}$.
    
    \section{Computational details and results} \label{sec:compdetails}

    \def \arraystretch{1.5} 
    \setlength{\tabcolsep}{.65em} 
    \begin{table*} 
    \centering 
    \begin{tabular} 
    {cccccc} 
    \toprule 
    Molecule & Modes & Modals & Toffoli gates & Qubits & Electronic structure\\
    \midrule 
    Hydrogen peroxide (3M) & 6 & 5 & $\num{2.94e+12}$ & 430 & CCSD(F12*)(T*)/cc-pVTZ-F12 \\ 
    Formaldehyde (2M) & 6 & 5 & $\num{2.00e+10}$ & 416 & CCSD(T)/cc-pVTZ \\ 
    Formaldehyde (3M) & 6 & 5 & $\num{5.05e+11}$ & 442 & CCSD(T)/cc-pVTZ \\ 
    PAH1 (2M) & 30 & 3 & $\num{9.44e+09}$ & 1765 & GFN2-xTB\\ 
    PAH1 (3M) & 30 & 5 & $\num{2.73e+13}$ & 2285 & GFN2-xTB\\ 
    PAH2 (2M) & 48 & 3 & $\num{2.17e+10}$ & 2859 & GFN2-xTB\\ 
    PAH2 (3M) & 48 & 5 & $\num{5.27e+14}$ & 3975 & GFN2-xTB\\ 
    PAH3 (2M) & 66 & 3 & $\num{4.75e+10}$ & 4123 & GFN2-xTB\\ 
    PAH4 (2M) & 84 & 3 & $\num{5.51e+10}$ & 5243 & GFN2-xTB\\ 
    PAH5 (2M) & 102 & 3 & $\num{8.19e+10}$ & 6361 & GFN2-xTB\\ 
    PAH6 (2M) & 120 & 3 & $\num{1.11e+11}$ & 7479 & GFN2-xTB\\ 
    PAH7 (2M) & 138 & 3 & $\num{1.41e+11}$ & 9024 & GFN2-xTB\\ 
    PAH8 (2M) & 156 & 3 & $\num{2.01e+11}$ & 10191 & GFN2-xTB\\ 
    \bottomrule 
    \end{tabular} 
    \caption{QPE costs and computational details for all benchmark molecules using chemically accurate decomposed vibrational Hamiltonians using three modals per mode for all molecules. The decomposed vibrational Hamiltonians are truncated to ensure chemical accuracy and parallelization and grouping methods are utilized to reduce circuit depths.}
    \label{tab:appendix_qpe_costs} 
    \end{table*} 
    
    The \ac{qpe} costs and computational details for all benchmark molecules are presented in \cref{tab:appendix_qpe_costs}. The molecular geometry of each compound was optimised using later specified electronic structure programs and vibrational coordinates were generated using the MidasCpp program~\cite{christiansen_midascpp_nodate}. These coordinates were, unless otherwise specified, normal mode coordinates based on the Hessian computed either with the electronic structure theory program or numerically using MidasCpp. Lastly, the \ac{adga}~\cite{sparta_adaptive_2009}, implemented in MidasCpp, was used to construct the \ac{pes} in the region of the equilibrium geometry with \ac{vscf} calculations carried out in large B-spline bases. All electronic structure single point calculations used in the \ac{pes} construction were computed with the same program as the geometry optimisation.

    \begin{figure*}
        \centering
        \resizebox{\textwidth}{!}{
        \begin{tikzpicture}
            \node[anchor=center,inner sep=0] at (0,0)
            { \chemname[1.5ex]{ \chemfig{*6(-=-=-=)}}{\mediumlarge PAH1} };
            \node[anchor=center,inner sep=0] at (3.1,0)
            { \chemname[1.5ex]{ \chemfig{*6(-=*6(-=-=-)-=-=)}}{\mediumlarge PAH2} };
            \node[anchor=center,inner sep=0] at (7.9,0)
            { \chemname[1.5ex]{ \chemfig{*6(-=*6(-=*6(-=-=-)-=-)-=-=)}}{\mediumlarge PAH3} };
            \node[anchor=center,inner sep=0] at (14.35,0)
            { \chemname[1.5ex]{ \chemfig{*6(-=*6(-=*6(-=*6(-=-=-)-=-)-=-)-=-=)}}{\mediumlarge PAH4} };
            \node[anchor=center,inner sep=0] at (7.91,-3)
            { \chemname[1.5ex]{ \chemfig{*6(-=*6(-=*6(-=*6(-=*6(-=*6(-=*6(-=*6(-=-=-)-=-)-=-)-=-)-=-)-=-)-=-)-=-=)}}{\mediumlarge PAH8} };
        \end{tikzpicture}
        }
        \begin{tikzpicture}
            \node[anchor=center,inner sep=0] at (-3,0)
            { \chemname[1ex]{\chemfig{ [:90] H-[1]C(=O)-[-1]H }}{Formaldehyde} };

            \node[anchor=center,inner sep=0] at (3,0)
            { \chemname[1ex]{\chemfig{ H-[1]O=O-[1]H }}{Hydrogen peroxide}  };
        \end{tikzpicture}
        \caption{Illustration of selected \aclp{pah} in the \ac{pah} test set, formaldehyde, and hydrogen peroxide.}
        \label{fig:pah_molecules}
    \end{figure*}
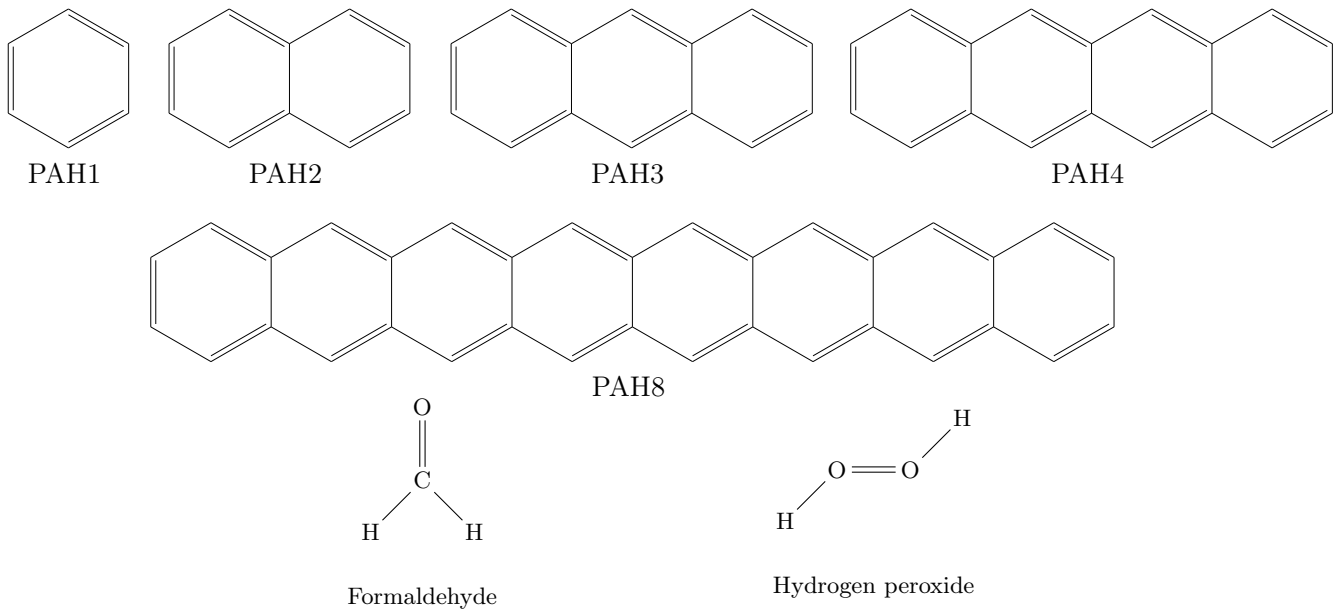

    \subsection{\Aclp{pah}} \label{sec:pah_pes}

    The \acl{pah} (\acs{pah}) test set consists of a linear chain of benzene rings as illustrated in \cref{fig:pah_molecules}. For each compound the electronic structure calculations for geometry optimisation and \ac{pes} construction were done at the GFN2-xTB~\cite{bannwarth_gfn2-xtbaccurate_2019} level of theory using xTB~\cite{bannwarth_extended_2021}. These surfaces were computed using both normal coordinates based on the semi-numerical Hessian from xTB.

    \subsection{Formaldehyde and hydrogen peroxide}

    For formaldehyde, the \ac{pes} of the one-, two- and three-mode \ac{pes} were constructed at the CCSD(T)/cc-pVTZ level of theory. For hydrogen peroxide, the \ac{pes} of the one-, two- and three-mode \ac{pes} were constructed at the CCSD(F12*)(T*)/cc-pVTZ-F12 level of theory. The electronic structure calculations were performed using TURBOMOLE~\cite{balasubramani_turbomole_2020}.

% OLD SCHOOL BBL INCLUDE 

\end{document}